\documentclass[authorversion]{acmart}
\usepackage[utf8]{inputenc}
\usepackage{array}
\usepackage{graphicx}
\usepackage{multirow}
\usepackage{url}
\usepackage{pdfpages}
\usepackage{booktabs}

\title{Who Shares What? An Empirical Analysis of Security Conference Content Across Academia and Industry}
\author{Lukas Walter, Clemens Sauerwein, and Daniel W. Woods}


\begin{document}
\setcopyright{cc}
\setcctype{by}
\acmJournal{DTRAP}
\acmYear{2026} \acmVolume{1} \acmNumber{1} \acmArticle{}
\acmMonth{1} \acmPrice{} \acmDOI{10.1145/3788676}

\begin{abstract}
Security conferences are important venues for information sharing, where academics and practitioners share knowledge about new attacks and state-of-the-art defenses. 
Despite their importance, researchers have not systematically examined who shares information and which security topics are discussed.
To address this gap, our paper characterizes the speakers, sponsors, and topics presented at prestigious academic and industry security conferences. 
We compile a longitudinal dataset containing 9,728 abstracts and 1,686 sponsors across four academic and six industry conferences.
Our findings show limited information sharing between industry and academia. Conferences vary significantly in how equitably talks and authorship are distributed across individuals.
The topics of academic and industry abstracts display consistent coverage of techniques within the MITRE ATT\&CK framework.
Top-tier academic conferences, as well as DEFCON and Black Hat, address the governance, response, and recovery functions of the NIST Cybersecurity Framework inconsistently.
Commercial information security and insurance conferences (RSA, Gartner, Advisen and NetDiligence) more consistently cover the framework.
Prevention and detection were the most common topics in the sample period, with no clear temporal trends.
\end{abstract}

\authorsaddresses{\textbf{Lukas Walter}: University of Innsbruck, Austria; E-mail: lukwalter7@gmail.com\\ 
\textbf{Clemens Sauerwein}: University of Innsbruck, Austria; E-mail: clemens.sauerwein@uibk.ac.at\\
\textbf{Daniel W. Woods}: University of Edinburgh, United Kingdom; E-mail: daniel.woods@ed.ac.uk}
\maketitle

\section{Introduction}
\label{sec:introduction}
Information security knowledge is produced by both academic and industry researchers.
However, these two communities differ in their norms and values, including what constitutes ``rigorous'' research and the role of peer review.
Despite these differences, both academic and industry researchers value the exchange of knowledge that occurs when experts convene to present and discuss findings at conferences.

For many professionals in both academia and industry, these events represent important venues for professional development and community engagement.
Attendees often travel internationally and spend significant amounts on registration fees and related expenses.
Although some of these activities are not directly related to formal knowledge exchange (e.g., hallway discussions and informal extracurricular activities), most participants devote time to listening to presentations that convey insights into novel attacks, mitigation techniques, lessons learned, and emerging security concepts.

Despite the substantial investment of time and resources by attendees and their employers, security conferences have not been systematically studied as part of the security information sharing ecosystem~\cite{skopik2016problem, laube2017strategic, ainslie2023cyber}.
To our knowledge, no prior research has analyzed the speakers or topics featured at industry conferences.
Existing studies on academic security conferences have primarily focused on bibliometrics~\cite{wendzel2020not, balzarot2022} and evaluation methodologies~\cite{carver2016establishing, burcham2017characterizing}.
This research gap is significant because conferences influence which security topics receive attention, thereby shaping the direction of research and practice.
While conferences provide a platform for selected researchers to present their work, low acceptance rates necessarily exclude many others.\footnote{The acceptance rates at top-tier academic conferences like IEEE Security and Privacy, USENIX Security, ACM CCS, and NDSS range from 5--20\% depending on the year. Industry conferences like Black Hat, DEFCON and RSA do not report acceptance rates.} \\
To address the research gap, our study characterizes information sharing at leading academic and industry security conferences by posing the following research questions: 
\begin{itemize} \setlength{\itemindent}{0.6em}  
 \setlength\itemsep{0em}
    \item[\textbf{RQ1}:] Who speaks at security conferences?
    \item[\textbf{RQ2}:] About which topics?
    \item[\textbf{RQ3}:] How does this vary by conference and over time?
\end{itemize}
We collected a dataset containing titles, abstracts, and author or speaker information from four academic and six industry conferences held between 2014 and 2022. 
To analyze thematic content, we classified abstracts as offensive, defensive, or neutral (for theoretical work). Offensive talks were mapped to the MITRE ATT\&CK Framework\footnote{\url{https://attack.mitre.org/} (accessed  19 October 2025)}, and defensive talks to the NIST Cybersecurity Framework\footnote{\url{https://www.nist.gov/cyberframework} (accessed 19 October 2025)}. 

Our contributions are as follows:
\begin{itemize}  \setlength{\itemindent}{0.8em}  \setlength{\itemsep}{0em}
    \item[\textbf{RQ1:}] Among academic conferences, NDSS exhibits the most equitable distribution of speakers, whereas ACM CCS shows the least. Cross-participation between academic, industry, and insurance conferences is rare. We found no evidence that sponsorship significantly affects speaker selection at industry conferences.
    \item[\textbf{RQ2:}] Defensive talks are more common than offensive talks at all conferences except DEFCON. Defensive talks focus on protective technology, detection, and response. Offensive talks are more evenly distributed across offensive security categories in MITRE ATT\&CK.
    \item[\textbf{RQ3:}] The distribution of topics remains relatively stable over time. Compared among conferences, DEFCON covers similar topics to academic conferences. RSA and Gartner emphasize on governance, while cyber insurance conferences focus on incident response.
    \item[\textbf{Data:}] The dataset compiled for this study was made publicly available to support replication and further analysis \cite{walter_2024_15989593}.
    \end{itemize}
The remainder of this paper is structured as follows. Section~\ref{sec:related_work} discusses related work in the field of information sharing. Section~\ref{sec:methodology} describes the research methodology. 
Section~\ref{sec:results} presents the empirical results. Section~\ref{sec:limitations} discusses the limitations of the study. Section~\ref{sec:discussion} outlines the implications of our findings and provides recommendations. 
Section~\ref{sec:conclusion} concludes the paper.

\section{Related Work} 
\label{sec:related_work}
Defenders use and share various forms of security information to better calibrate their defensive posture to the threat landscape~\cite{webb2014situation, ahmad2021can}.
Most research focuses on sharing real-time technical indicators~\cite{laube2017strategic}, which we review in Section~\ref{subsec:infosharing}.
Other communication channels through which defenders acquire and disseminate security knowledge are discussed in Section~\ref{subsec:otherchannels}.

\subsection{Information Sharing}
\label{subsec:infosharing}
The two most common forms of security information sharing are threat intelligence (TI) and vulnerability management. A range of schema and standards have been introduced to facilitate sharing~\cite{kampanakis2014security}, including MAEC, OVAL, XCCDF, CPE, CVE, and CVSS.
Governments have passed regulations to support information sharing~\cite{yang2020impact}, funded Computer Emergency Response Teams (CERTs) to coordinate vulnerability disclosure~\cite{slayton2020trusting}, and directly shared threat intelligence\footnote{\url{https://www.cisa.gov/known-exploited-vulnerabilities-catalog} (accessed 19 October 2025)}.
Commercial providers also offer threat intelligence and surrounding services~\cite{wagner2019cyber,sauerwein2017threat,brown2015cyber,sauerwein2019analysis}.
The associated platforms provide information enrichment and analysis functions to obtain targeted and actionable threat intelligence~\cite{sauerwein2021threat,dandurand2013towards,brown2015cyber,chismon2015threat}.

Recent empirical studies have questioned the quality of threat intelligence.
Li et al.~\cite{li2019reading} find little overlap between different TI feeds and cannot identify a way to distinguish quality, which confirms findings from other studies~\cite{thomas2016abuse, bouwman2022helping}.
Expert interviews with TI practitioners reveal processes to evaluate quality are informal~\cite{bouwman2020different}, which is to be expected given this remains an open research challenge~\cite{zibak2022threat}.
Encouragingly, a study of threat intelligence during COVID-19 found evidence that collaboration, in the form of aggregating commercial TI feeds, improved coverage of threats~\cite{bouwman2022helping}.

In the area of vulnerability management, bug bounty programs have proliferated, run both independently by vendors~\cite{zhao2017devising} and on platforms like HackerOne~\cite{sridhar2021hacking}.
Empirical evaluations support the efficacy of bug bounty programs over internal testing~\cite{finifter2013empirical, zhao2015empirical}, even though they follow similar processes~\cite{votipka2018hackers}. Recent work has shown bug hunters in China organize into teams, in part to facilitate information sharing~\cite{piao2025study}. However, many firms do not patch even when the vulnerability is known and the patch is available~\cite{frei2006large, li2017large}. This has motivated research into understanding expert users who apply patches~\cite{li2019keepers, mathur2016they}, and notifications that nudge firms towards patching~\cite{stock2016hey, li2016you, ccetin2019tell}.

The sheer volume of information shared about security vulnerabilities has created a prioritization challenge, making it difficult for defenders to identify which issues merit attention~\cite{jacobs2020improving, de2023no}.
Information overload similarly affects threat intelligence~\cite{bouwman2020different}. 
This motivates research exploring online advice and knowledge sharing that may be easier for the recipients to digest than technical indicators.


\subsection{Advice and Knowledge Sharing} \label{subsec:otherchannels}
Online forums are often used to solve security and privacy problems~\cite{li2023itsup, tahaei2020understanding}, including by developers~\cite{acar2016you}. 
In some cases, this involves copying insecure code snippets directly from the forum post~\cite{fischer2017stack}. 
This has motivated research into how to nudge developers towards better choices~\cite{fischer2019stack}.

Online security advice targeted at the general public is also widely available~\cite{redmiles2016think}. An analysis of over a thousand sources of advice reveals that the specific recommendations are actionable and comprehensible, albeit with room for improvement~\cite{redmiles2020comprehensive}. However, the sources recommended 374 unique behaviors, which leads to information overload~\cite{redmiles2020comprehensive}.
Some users address the quality issue by looking to trusted sources~\cite{redmiles2016think}.

Conferences address this issue by curating the agenda, via peer review in academia. Research into what information is shared at these conferences has been small-scale. For example, Carver et al.~\cite{carver2016establishing} identify which evaluation method (e.g., proof or empirical) was used by each paper in the 2015 IEEE Security and Privacy proceedings. This study was extended to cover the 2016 ACM CCS proceedings~\cite{burcham2017characterizing}. A bibliometric analysis was conducted of around 30,000 academic computer security papers~\cite{wendzel2020not}, focusing on statistics like the number of words, authors, and citations in each paper.

We are not aware of analysis of industry conferences, even though practitioner conferences have tens of thousands of attendees (e.g.\,45k at RSA).
Across both academia and industry, it remains unclear who speaks/publishes at security conferences (\textbf{RQ1}), about which topics (\textbf{RQ2}), and how this changes over time and conferences (\textbf{RQ3}).

\section{Methodology} 
\label{sec:methodology}

We focused on prestigious security conferences, as they exerted the greatest influence within both academia and industry. Each talk can be analyzed as a video recording, via the associated paper, or metadata such as the abstract and title. While full academic papers contained the most comprehensive information, their industry equivalents (e.g., white papers) were not consistently available. Similarly, video recordings of talks were inconsistently published, and automated transcription remained imperfect. For these reasons, we relied on abstracts and titles, which were consistently available across both domains.

\begin{figure}
    \centering
    \includegraphics[width=0.8\textwidth]{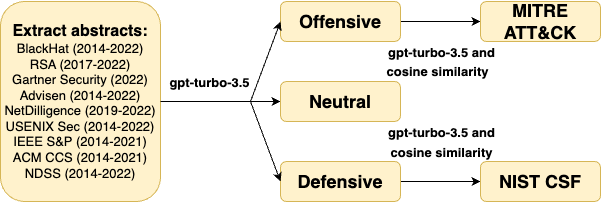}
    \caption{Two-stage classification process. In the first stage, GPT-based natural language processing is used to classify abstracts. In the second stage, both GPT and cosine similarity are applied to map abstracts to cybersecurity frameworks.}
    \label{fig:workflow}
\end{figure}

Section~\ref{subsec:datacollection} describes our corpus of conference abstracts. Section~\ref{subsec:analysis} explains the natural language processing (NLP) techniques used to classify talks. Section~\ref{subsec:validation} outlines our validation strategy. Section~\ref{subsec:ethics} discusses ethical considerations and open science. Figure~\ref{fig:workflow} provides a high-level overview of our research design.

\subsection{Data Collection} \label{subsec:datacollection}

\paragraph{\textbf{Conferences}}
For academia, we focused on top-tier computer security conferences~\cite{soneji2022flawed}: USENIX Security, IEEE S\&P, ACM CCS, and NDSS. 
Cryptography-specific venues were excluded because their focus did not align with the frameworks considered (see Section~\ref{subsec:analysis}). 
Journals were also excluded, as the absence of in-person presentations undermines the comparability with industry conferences.

For industry, we conducted a structured web search for articles featuring the `top security conferences'. 
Only Black Hat, DEFCON and RSA were consistently included across all identified lists. For example, the SANS Institute hosts influential topic-specific conferences (e.g.\,Threat Intelligence or Network Security), but no specific event is consistently included as the flagship event. We consider these seven conferences to be the most prestigious computer security conferences.

To diversify our sample, we included Gartner Security - given its influence over procurement decisions - and two leading cyber insurance conferences (Advisen and NetDiligence). Cyber insurance is increasingly used by firms to manage cybersecurity risk~\cite{dambra2020sok}. Its inclusion offers an informative comparison, particularly because the insurance community’s roots lay in finance rather than IT. Table~\ref{tab:conferences_metadata} displays the conferences in our sample, along with some simple statistics.
\begin{table*}
\centering \footnotesize
\scalebox{1}{
\begin{tabular}{llccccccc}
\toprule
\textbf{Conference} & \textbf{Sample Window} & \textbf{Type} & \multicolumn{2}{c}{\textbf{Talks per year}} & \multicolumn{2}{c}{\textbf{Characters in title}} & \multicolumn{2}{c}{\textbf{Words per abstract}} \\
& & & \textbf{$\mu$} & \textbf{$\sigma$} & \textbf{$\mu$} & \textbf{$\sigma$} & \textbf{$\mu$} & \textbf{$\sigma$} \\
\midrule
Black Hat & 2014 - 2022 & industry (InfoSec) & 252 & 53.83 & 54.82 & 26.43 & 170.98 & 99.16 \\
RSA & 2017 - 2022 & industry (InfoSec) & 338.33 & 34.98 & 55.69 & 17.18 & 75.75 & 26.98 \\
DEFCON & 2014 - 2022 & industry (InfoSec) & 92.22 & 32.19 & 56.10 & 25.75 & 176.72 & 77.72 \\
Gartner Security & 2022 & industry (InfoSec) & 321.0 & 0.0 & 68.07 & 19.38 & 66.85 & 20.38 \\
Advisen & 2014 - 2022 & industry (Insurance) & 10.0 & 4.92 & 38.81 & 17.62 & 58.67 & 22.25 \\
NetDiligence & 2019 - 2022 & industry (Insurance) & 25.0 & 6.22 & 51.50 & 20.35 & 36.76 & 27.36 \\
\midrule
USENIX Security & 2014 - 2022 & academic & 132.78 & 68.33 & 72.30 & 22.40 & 184.12 & 62.10 \\
IEEE S\&P & 2014 - 2022 & academic & 82.22 & 34.08 & 69.77 & 21.42 & 222.52 & 63.53 \\
ACM CCS & 2014 - 2022 & academic & 166.33 & 34.47 & 67.99 & 21.44 & 186.21 & 82.81 \\
NDSS & 2014 - 2022 & academic & 73.0 & 15.80 & 72.11 & 19.79 & 186.54 & 82.22 \\
\bottomrule
\end{tabular}
}
\caption{Conference metadata. Descriptive statistics showing the mean ($\mu$) and standard deviation ($\sigma$) across all years for the number of talks per year, title length in characters, and abstract length.}
\label{tab:conferences_metadata}
\end{table*}

\begin{table*}
\centering \footnotesize
\scalebox{1}{
\begin{tabular}{lccc}
\toprule
\textbf{Conference} & \multicolumn{2}{c}{\textbf{Sponsors}} & \\
& \textbf{$\mu$} & \textbf{$\sigma$} & \textbf{Notes} \\
\midrule
Black Hat & 179.0 & 54.29 & \\
RSA & 75.0 & 0.0 & contains data only from year 2022. \\
DEFCON & 0 & 0 & \\
Gartner Security & N/A & N/A & sponsor data was not available\\
Advisen & 18.13 & 5.95 & year 2020 is missing. \\
NetDiligence & 56.50 & 15.32 & \\
\midrule
USENIX Security & 15.22 & 3.33 & \\
IEEE S\&P & 14.67 & 6.22 & \\
ACM CCS & 13.87 & 3.33 & year 2018 is missing. \\
NDSS & 10.14 & 3.27 & only years from 2016 - 2022\\
\bottomrule
\end{tabular}
}
\caption{Sponsorship data. Mean ($\mu$) and standard deviation ($\sigma$) of the number of sponsors per conference. Notes indicate data-specific limitations.}
\label{tab:conferences_sponsorship}
\end{table*}

\begin{table*}
\centering \footnotesize
\scalebox{1}{
\begin{tabular}{lccc}
\toprule
\textbf{Conference} & \multicolumn{2}{c}{\textbf{Gini Index}} \\
& \textbf{1} & \textbf{1/N} \\
\midrule
Black Hat & 0.234 & 0.381 \\
RSA & 0.288 & 0.441 \\
DEFCON & 0.179 & 0.382 \\
Gartner Security & 0.328 & 0.427 \\
Advisen/Zywave & 0.244 & 0.383 \\
NetDiligence & 0.159 & 0.269 \\
\midrule
USENIX Security & 0.284 & 0.417 \\
IEEE S\&P & 0.280 & 0.380 \\
ACM CCS & 0.325 & 0.451 \\
NDSS & 0.209 & 0.376 \\
\bottomrule
\end{tabular}
}
\caption{Speaker distribution equality. The Gini index is calculated for the total number of talks per individual across all years. The two columns report Gini index values under different normalization factors (1 and 1/N).}
\label{tab:conferences_gini}
\end{table*}

\paragraph{\textbf{Data Extraction}}
The dataset used in this study was collected exclusively from publicly available sources, without reliance on institutional subscriptions or paywalled materials. Data was retrieved directly from official conference websites or openly accessible conference proceedings using web content mining techniques. We extracted freely available metadata, including titles, author names, abstracts, and sponsor lists, as published on conference or publisher landing pages. Consequently, no special permissions were required from conference organizers or authors. A semi-automated process was employed to extract information on speakers, affiliations, titles, abstracts, and dates. For academic venues, we included keynotes and paper presentations for which an abstract was available, excluding poster sessions. For industry conferences, we considered all talks with an abstract listed in the conference agenda and omitted non-technical sessions such as networking events or welcome remarks.
The resulting dataset comprises 9,728 talks (5,639 industry; 4,089 academic) and 1,686 sponsor entries, of which 451 corresponded to academic and 371 to cyber insurance conferences.

\subsection{Data Analysis}  
\label{subsec:analysis}
The analysis for \textbf{RQ1} involved calculating descriptive statistics about the distribution of talks, speakers, and sponsors.
To address \textbf{RQ2}, we applied natural language processing (NLP) methods to classify abstracts according to established cybersecurity frameworks (Figure~\ref{fig:workflow}). 
This deductive approach mapped textual content to structured taxonomies of offensive and defensive security practices. Two key decisions guided this analysis: (i) the selection of \emph{frameworks} and (ii) the choice of \emph{NLP techniques}.

\textbf{Choice of Framework} The chosen frameworks would ideally be: (i) comprehensive, (ii) widely recognized in practice, and (iii) freely accessible. The first criterion ensures us to meaningfully map more abstracts to the framework, which would not be possible with a digital forensics standard. The second criterion makes our results easier to understand, since some readers will always be familiar with the categories in the framework. The final criterion enables others to replicate our work.

For offensive frameworks, we considered MITRE ATT\&CK and Cyber Kill Chain framework. ATT\&CK framework offers greater depth, satisfying the first criterion.
For defensive frameworks, we evaluated NIST Cybersecurity Framework (CSF), ISO 27001 and CIS Critical Security Controls. We excluded ISO 27001 due to licensing costs. CIS standard was excluded because NIST CSF is broader, including governance. 
Although the choice of framework is important, mappings between frameworks are widely available\footnote{For example, CIS Controls v8 mapped to NIST CSF is available here: \url{https://www.cisecurity.org/insights/white-papers/cis-controls-v8-mapping-to-nist-csf} (accessed 19 October 2025)}. 

\textbf{NLP Techniques} We implemented a two-stage classification process. For the first classification phase, we used OpenAI's gpt-3.5-turbo model to classify abstracts as offensive, defensive, or neutral using the following minimal zero-shot prompt:
\begin{quote}
    ``Is this text cybersecurity offensive or defensive (just answer one word, namely: offensive, defensive or neutral)? $<$abstract$>$''
\end{quote}
Minimal prompting reduced output variation and improved scalability, as more detailed prompts in pilot tests led to longer, less consistent, and more ambiguous responses. 
Neutral talks, which comprise a minority (17\%) of the corpus, are not further classified, as they can not be clearly mapped to either frameworks (see Table~\ref{tab:crosssimilarity}). They are mentioned solely for completeness.

For the second stage of the classification, we referred to the official documentation of each standard to establish themes and sub-themes. For MITRE ATT\&CK Framework, we used the 14 tactics as themes and the 196 techniques as sub-themes. For NIST Cybersecurity Framework, we used the 22 categories as themes and the 98 sub-categories as sub-themes. We estimated the similarity between each abstract and the corresponding (sub) theme using two NLP techniques: cosine similarity and a Large Language Model (LLM).

To calculate \emph{cosine similarity}, we preprocessed text by removing stop words (NLTK stopword list) and lemmatized terms using WordNetLemmatizer. Texts were represented as embeddings via SentenceTransformer (stsb-mpnet-base-v2) and compared to framework categories using cosine similarity. This process was applied to both NIST CSF and MITRE ATT\&CK hierarchies.

For the \emph{Large Language Model}, we used OpenAI's gpt-3.5-turbo model using the following zero-shot prompt:
\begin{quote}
    ``What are the MITRE ATT\&CK tactics that can be identified in the following text. Only return the tactics which are identified: 
 $<$abstract$>$''
\end{quote}
A similar prompt was used for the NIST CSF categories. 
Each theme was queried individually, producing binary (0/1) outputs per talk. To calculate aggregate similarity to each theme, we took the average across all talks in a specific sub-sample.

\subsection{Validation} \label{subsec:validation}
Validation was difficult because there is no existing ground truth mapping for abstract talks to cybersecurity frameworks. Further, frameworks are high-level with blurry boundaries between categories. 
We employed a three-pronged validation approach to address these issues, consisting of: (i) human annotation, (ii) cross-method agreement, and (iii) theoretical consistency.



For human annotation, we manually labeled 264 abstracts and compared these to GPT classification.
The model achieved a $92.4\%$ concordance with the human annotators. Based on the selected sample ($n=264$) from the total population of $N=9,728$ abstracts, a $95\%$ confidence interval for true population concordance was determined at $\pm 3.16$ percentage points.
This calculation incorporated the Finite Population Correction (FPC) to accurately reflect the known, finite size of the total abstract population.

For the cross-method agreement, we compared the topic classifications across NLP techniques. 
Because cosine similarity produced continuous values, we applied a threshold to align the number of positive (1) and negative (0) classifications.
We set the threshold such that both GPT and cosine similarity would have the same number of 0s and 1s. Otherwise the scores could be perfectly aligned, but appear to disagree purely due to setting different thresholds. 
For the ATT\&CK and NIST Cybersecurity framework categories, the two techniques agreed for 67\% and 81\% of the abstracts respectively.
This ranged across individual categories, from a low of 52\% up to 95\% (see Table~\ref{fig:individualagreement}).
The categories with high accuracy are partly driven by the class imbalance as so few defensive abstracts cover \emph{Maintenance} (from NIST CSF) or \emph{Resource Development} (from MITRE ATT\&CK).

For the theoretical consistency, we examined whether offensive and defensive talks displayed similarity patterns. It might be expected that offensive or defensive talks display the highest similarity to the offensive or defensive frameworks, while neutral talks would exhibit low similarity to both frameworks. Table~\ref{tab:crosssimilarity} compares the performance of the two classification techniques in aligning with our labeling of abstracts as offensive, defensive, or neutral. Specifically, it presents the average similarity scores between each category of talk and the corresponding cybersecurity frameworks. The results show that the cosine similarity scores display the expected ranking: defensive talks exhibit the highest similarity to the NIST CSF (0.317), and offensive talks align most closely with the ATT\&CK framework (0.452). However, this is not true for OpenAI's gpt-3.5-turbo model, as offensive talks display higher similarity to both frameworks, and defensive talks display the lowest similarity to NIST CSF.

\begin{table}
    \centering \footnotesize \setlength{\tabcolsep}{2pt}
    \begin{tabular}{cccccccc} \hline
     
    \textbf{Talk} & \multicolumn{2}{c}{\textbf{Cosine Similarity}} & \multicolumn{2}{c}{\textbf{gpt-3.5-turbo}} \\
    \textbf{Category} & \textbf{NIST CSF} & \textbf{ATT\&CK} & \textbf{NIST CSF} & \textbf{ ATT\&CK} \\ \hline
    Offensive & 0.298 & 0.452 & 0.160 & 0.286\\
    Defensive & 0.317 & 0.411 & 0.135 & 0.236\\
    Neutral   & 0.266 & 0.342 & 0.149 & 0.155\\ \hline 
    \end{tabular}
    \caption{Average similarity between talk categories and cybersecurity frameworks, computed using GPT-based classification and cosine similarity techniques.}
    \label{tab:crosssimilarity}
\end{table}

\textbf{Summary}
These three validation tests indicates that our mapping captures meaningful signal, albeit with considerable noise - unsurprising given the subjective nature of the task.
The GPT method reliably classifies talks as offensive, defensive or neutral. However, the agreement between NLP techniques is disappointing.

Disagreement results from a mixture of errors/hallucinations and also differences in approach.
Cosine similarity focuses on keywords and phrases used in the descriptions of each framework.
This can lead to false positives when these words are used in a different context, while also leading to false negatives when abstracts use different words with the same meaning. 
On the other hand, the GPT method captures how words and phrases are used in the dataset, which is mainly user-submitted data from the Internet, which is vulnerable to context drift.

Going forward, we will report on aggregate results from OpenAI's gpt-3.5-turbo as it is a robust, widely used, and well-documented LLM variant. Aggregating the classifications helps mitigate random errors as these should cancel out. However, it does not mitigate systemic bias associated with GPT, which we discuss in Section~\ref{sec:limitations}. In the Appendix, we display supplementary figures based on an ensemble model, which outputs an equal weighting of the GPT and cosine similarity techniques.

\begin{table} \scriptsize 
\begin{tabular}{l@{}c|l@{}c}
\toprule								
ATT\&CK Tactic	&	Agree 	&	NIST CSF Function	&	Agree 	\\	\midrule
Credential Access	&	0.63	&	Asset Management	&	0.95	\\	
Execution	&	0.52	&	Business Environment	&	0.95	\\	
Impact	&	0.70	&	Governance	&	0.90	\\	
Persistence	&	0.56	&	Risk Assessment	&	0.84	\\	
Privilege Escalation	&	0.65	&	Risk Management Strategy	&	0.93	\\	
Lateral Movement	&	0.79	&	Supply Chain Risk Mgmt	&	0.93	\\	
Defense Evasion	&	0.57	&	Identity Mgmt and Access Control	&	0.86	\\	
Exfiltration	&	0.71	&	Awareness and Training	&	0.90	\\	
Discovery	&	0.52	&	Data Security	&	0.90	\\	
Collection	&	0.54	&	Info. Prot. Processes and Proc. &	0.94	\\	
Resource Dev. &	0.98	&	Maintenance	&	0.95	\\	
Reconnaissance	&	0.90	&	Protective Technology	&	0.58	\\	
Command \& Control	&	0.73	&	Anomalies and Events	&	0.65	\\	
Initial Access	&	0.56	&	Security Continuous Monitoring	&	0.84	\\	
	&		&	Detection Processes	&	0.62	\\	
	&		&	Response Planning	&	0.66	\\	
	&		&	Communications (Respond)	&	0.78	\\	
	&		&	Analysis	&	0.67	\\	
	&		&	Mitigation	&	0.67	\\	
	&		&	Improvements (Respond)	&	0.65	\\	
	&		&	Recovery Planning	&	0.79	\\	
	&		&	Improvements (Recover)	&	0.79	\\	
	&		&	Communications (Recover)	&	0.81	\\	\midrule
Mean	&	0.67	&	Mean	&	0.81	\\	\bottomrule
    \end{tabular}
\caption{Agreement between NLP techniques. Fraction of abstracts per category for which the GPT and cosine similarity classifications are consistent. \label{fig:individualagreement}}
    
\end{table}

\subsection{Open Science \& Ethics} \label{subsec:ethics}
We release our corpus of titles, abstracts, speakers, and sponsors in an open source repository, as well as our analysis scripts \cite{walter_2024_15989593}. 
We hope this allows other researchers to extend this research project. We discuss these directions in Section~\ref{sec:discussion}.

Our primary ethical consideration is the use of personal data—such as author names and affiliations—which we process and analyze. As this information is already publicly available and tied to individuals’ voluntary participation in public events, we believe the privacy risks are minimal. While obtaining affirmative consent from all speakers would be impractical, we acknowledge the concern and have taken care to handle the data responsibly. We believe the value of this work for open science justifies the approach, though future collaborations with conference organizers could further strengthen ethical alignment.

\section{Results}   
\label{sec:results}
This section presents the results of our analysis on the composition of talks and sponsors at each conference (Section~\ref{subsec:conference}), speakers (Section~\ref{subsec:speakers}), and abstract topics (Section~\ref{subsec:content}). 

\subsection{Conference Statistics}  
\label{subsec:conference}
\begin{figure}
    \centering
    \includegraphics[width=0.48\textwidth]{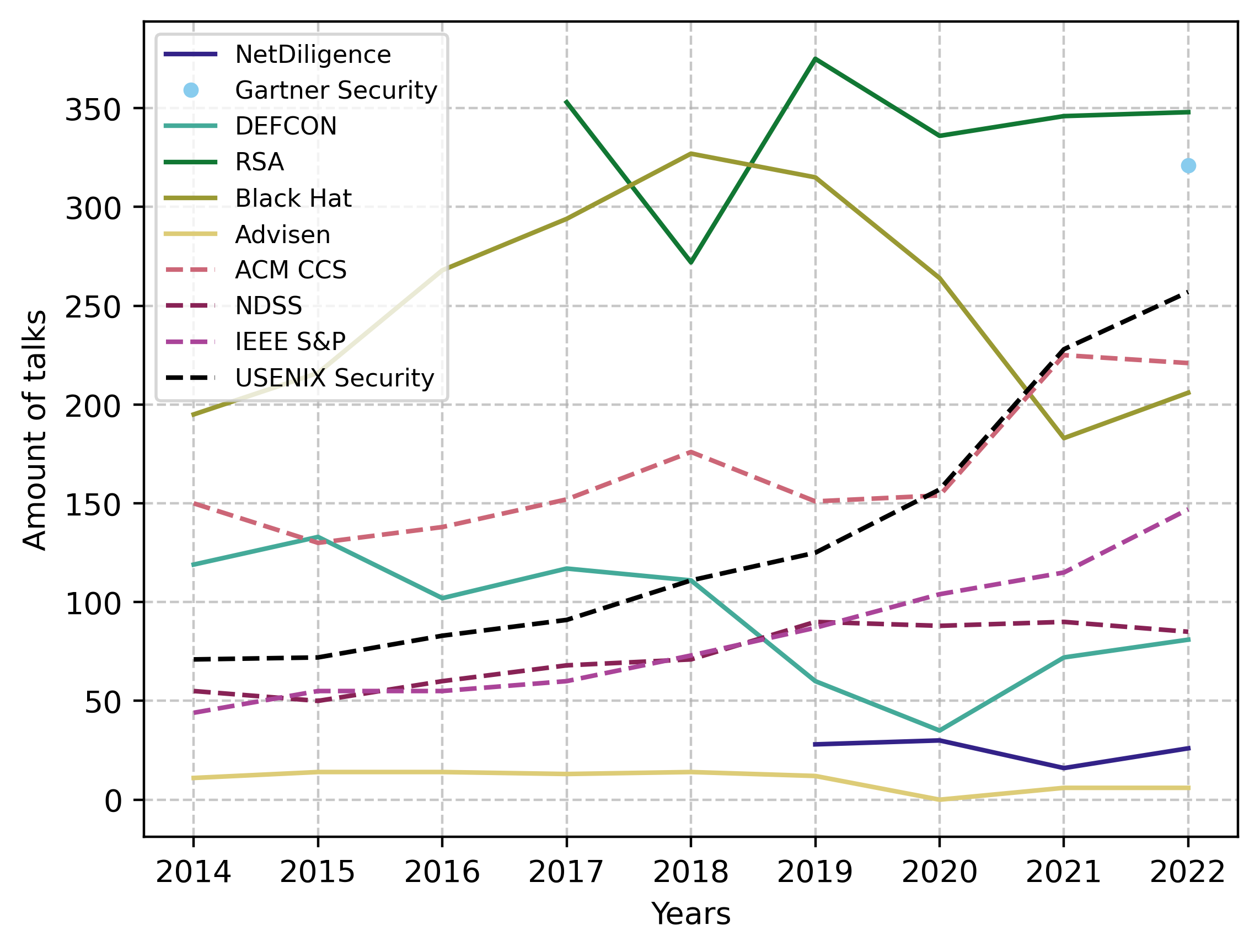}
    \caption{Number of talks presented at each conference from 2014 to 2022.}
    \label{fig:talksovertime}
\end{figure}

RSA and Black Hat had the most talks across the sample period (see Figure~\ref{fig:talksovertime}). 
Volatility in the size of Black Hat since 2019 is driven by the conference introducing/discontinuing sessions like Day Zero, CISO Summit, and Sponsored Workshops. 
The academic conferences display the highest rates of growth, as noted by Balzarotti~\cite{balzarot2022}. USENIX Security doubled in size from 2019 to 2021. The industry conferences do not display the same growth trend. 
Insurance conferences had the fewest talks, with the absolute number remaining stable over time.  This is perhaps surprising given the market grew by over 360\% between 2015 and 2021~\cite{naic2022report}.

\begin{figure*}[]{}
    \centering
    \includegraphics[width=1\textwidth]{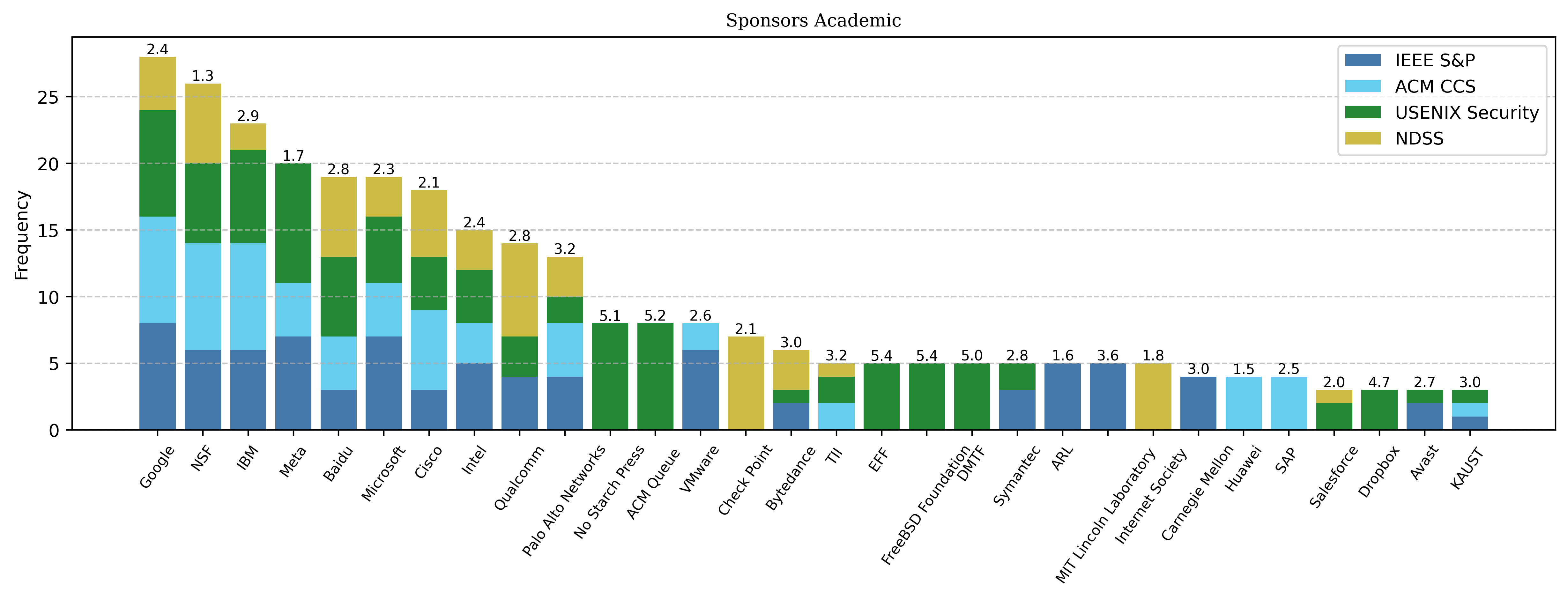}
    \caption{Top sponsors of academic conferences aggregated over time (2014–2022). Numbers above the bars indicate sponsorship tier of each sponsor.} 
    \label{fig:academicsponsorsfreq}
\end{figure*}

The academic conferences all had the longest abstracts in terms of word count. 
This should be considered an imperfect proxy for information content, especially given many conferences have a word limit.
Table~\ref{tab:conferences_metadata}  shows that IEEE S\&P displays the highest average at 223 words. 
Abstracts at Black Hat and DEFCON had a similar word count to ACM CCS, NDSS, and USENIX Security. 
These five conferences have a mean word count of between 170 and 190, which makes sense given original research is often presented at Black Hat and DEFCON. 
The conferences (RSA, NetDiligence, and Advisen) with the lowest word counts place less emphasis on research. 
There was less variance in the length of titles, both within and between conferences.

Black Hat has the most sponsors with an average of 197 per year (see Table~\ref{tab:conferences_sponsorship}). It is followed by RSA (75), NetDiligence (57), and Advisen (18). 
We were unable to find comprehensive sponsorship data for Gartner. The academic conferences all have between 10 and 16 sponsors per year. DEFCON does not accept sponsors. Conferences with fewer sponsors typically had longer abstracts.

\begin{figure}
    \centering
    \includegraphics[width=0.4\textwidth]{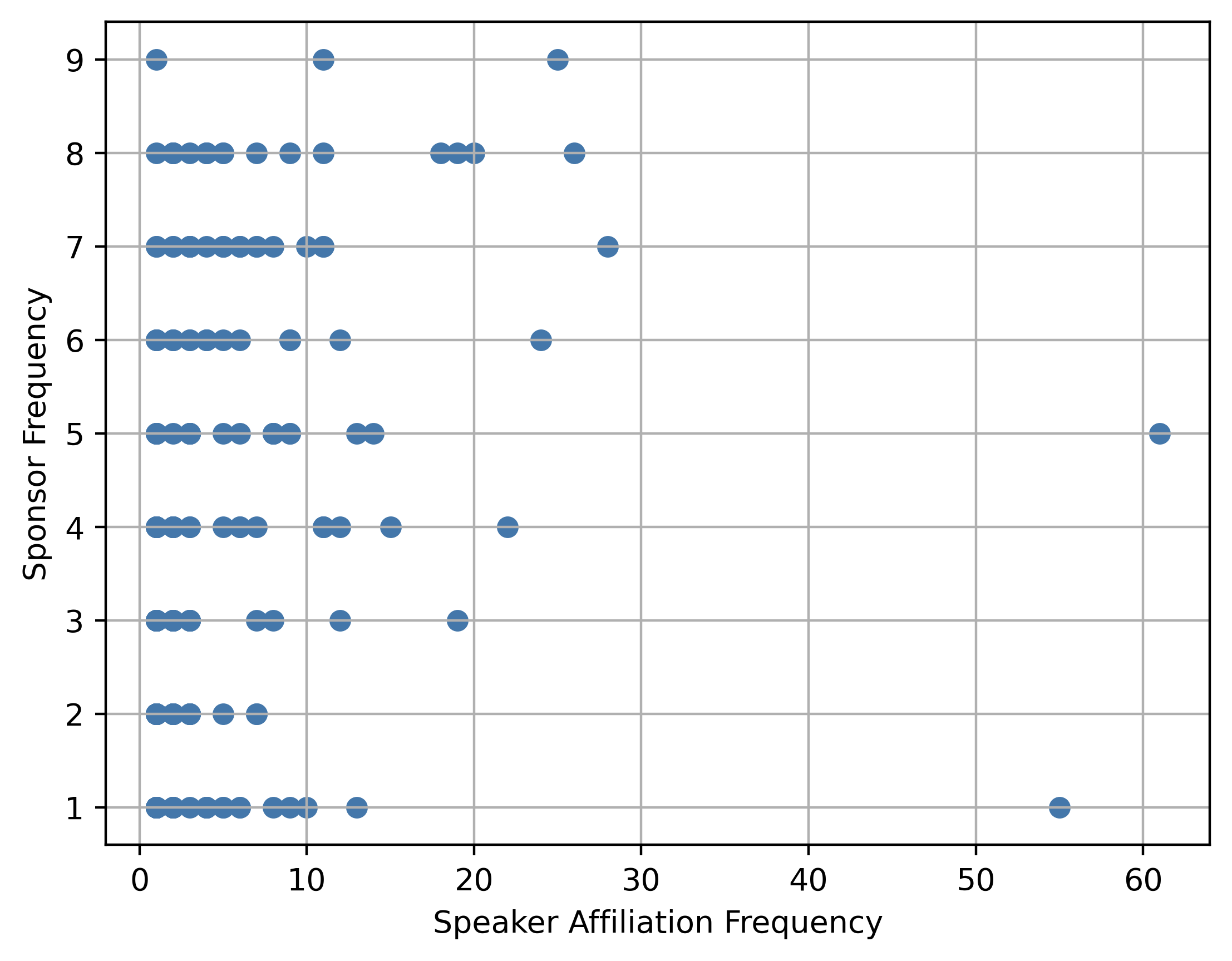}
    \caption{Relationship between speaker slots and sponsor slots at the Black Hat conference (2014–2022). Each point represents a company, with the x-axis showing speaker slots and the y-axis showing sponsor slots.}
    \label{fig:blackhat-sponsor-speaker-correlation}
\end{figure}

The academic conferences are sponsored by a mixture of public and private sources (see Figure~\ref{fig:academicsponsorsfreq}). Large global technology companies, along with the National Science Foundation, support all four conferences in a similar proportion. Information security (InfoSec) vendors tend to support specific conferences. Non-profits like the EFF, the FreeBSD Foundation, and DMTF only sponsor USENIX Security.

Although we could not collect information on the price of sponsorship, the numbers on top of the bars in Figure~\ref{fig:academicsponsorsfreq} represent the average sponsor tier of that company, where 1 is the highest.
This shows that the non-profits like EFF are consistently in lower tiers, while BigTech companies tend to be in the higher tiers.
However, the US-based National Science Foundation is most consistently a tier one sponsor.
No other country has an equivalent research institution that consistently sponsors these conferences.

The industry conferences are mainly sponsored by InfoSec vendors, with large technology companies sponsoring less frequently. 
This can be seen in Figure~\ref{fig:industrysponsorsfreq} in the Appendix. 
The analysis of sponsors of industry conferences is less comprehensive because of gaps in the data. 
The insurance conferences contain a wide range of sponsors (see Figure~\ref{fig:industrysponsorsfreq} in the Appendix). These include not just insurance companies, but also law firms, digital forensics providers, communications consultants, and data analytics firms. This is likely because these firms receive work from insurers~\cite{woods2023lessons}.

To probe whether sponsorship influences the number of talks, we analyze the only industry security conference for which we had longitudinal sponsor information, namely Black Hat.
Across all years, the conference had $592$ unique sponsors and $1493$ unique affiliation across speakers, which had an intersection of just $219$. This means that 85\% of the speakers' affiliated organizations never sponsored the conference, and 63\% of sponsors did not have an affiliated speaker at Black Hat. Figure~\ref{fig:blackhat-sponsor-speaker-correlation} shows the relationship between number of times an organization sponsored Black Hat and the number of affiliated talks. The two outliers are Microsoft and Google, who respectively had 61 and 55 affiliated talks.

\begin{figure*}
    \centering
    \includegraphics[width=\textwidth]{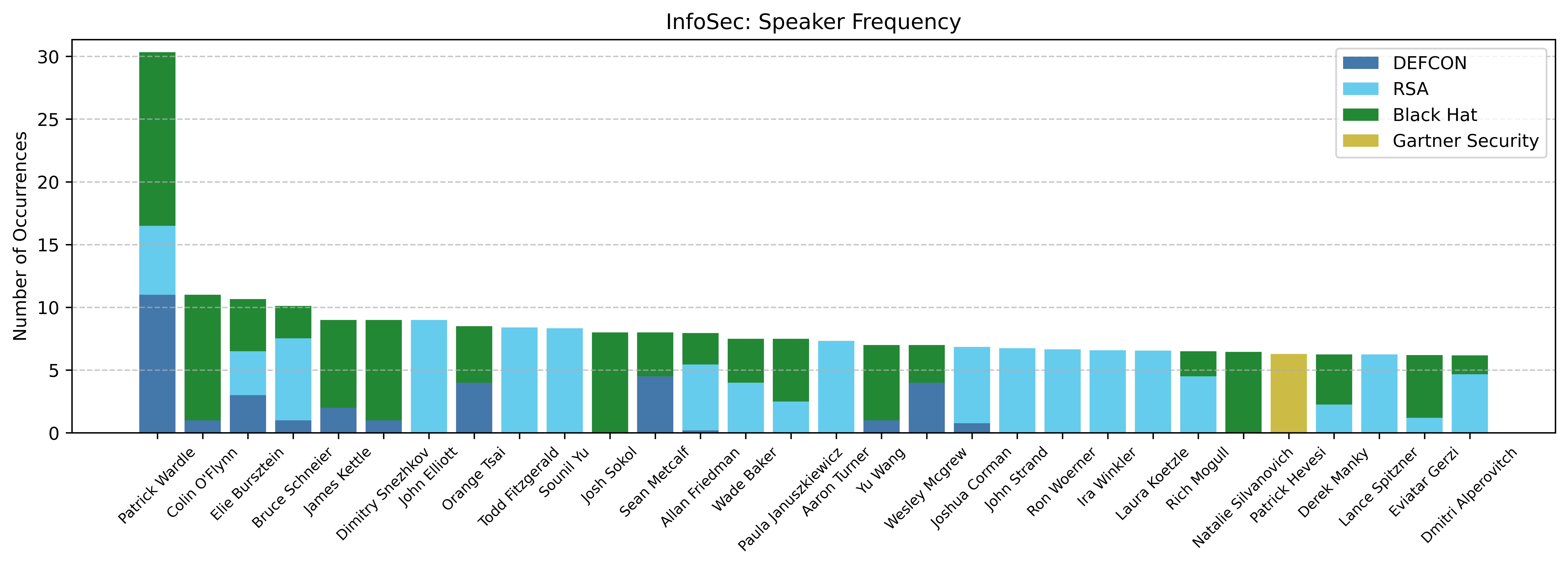}
    \includegraphics[width=\textwidth]{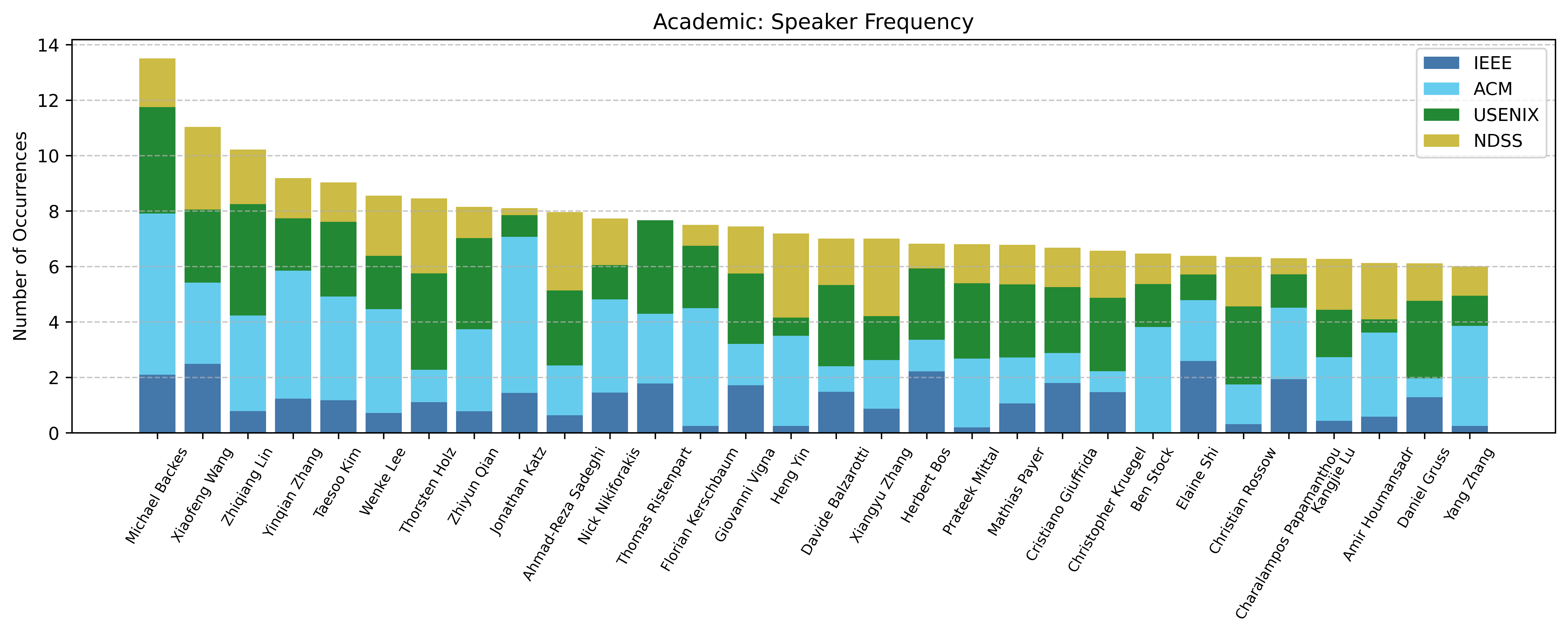}
    \includegraphics[width=\textwidth]{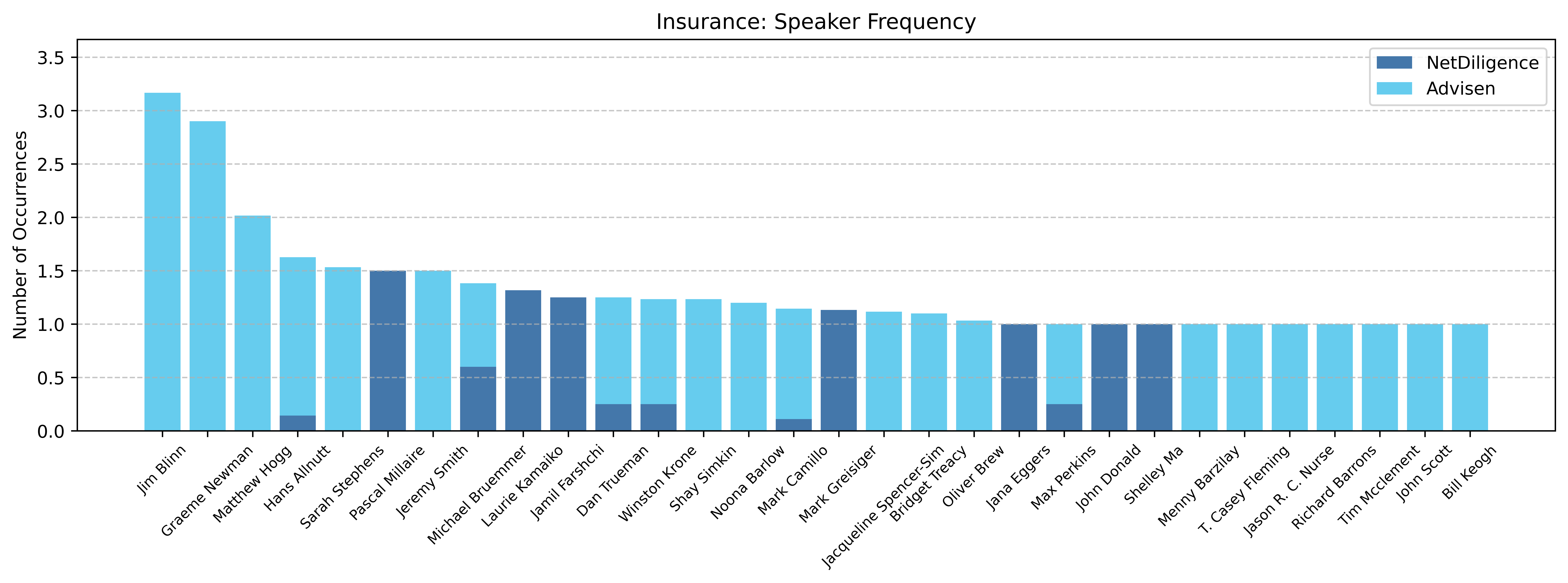}
    \caption{Number of talks per individual for the 30 most frequent speakers from 2014 to 2022 across industry, academic, and insurance conferences.} 
    \label{fig:freqanalysis}
\end{figure*}



\subsection{Speaker Statistics}
\label{subsec:speakers}
\begin{figure}
    \centering
    \includegraphics[width=0.4\textwidth]{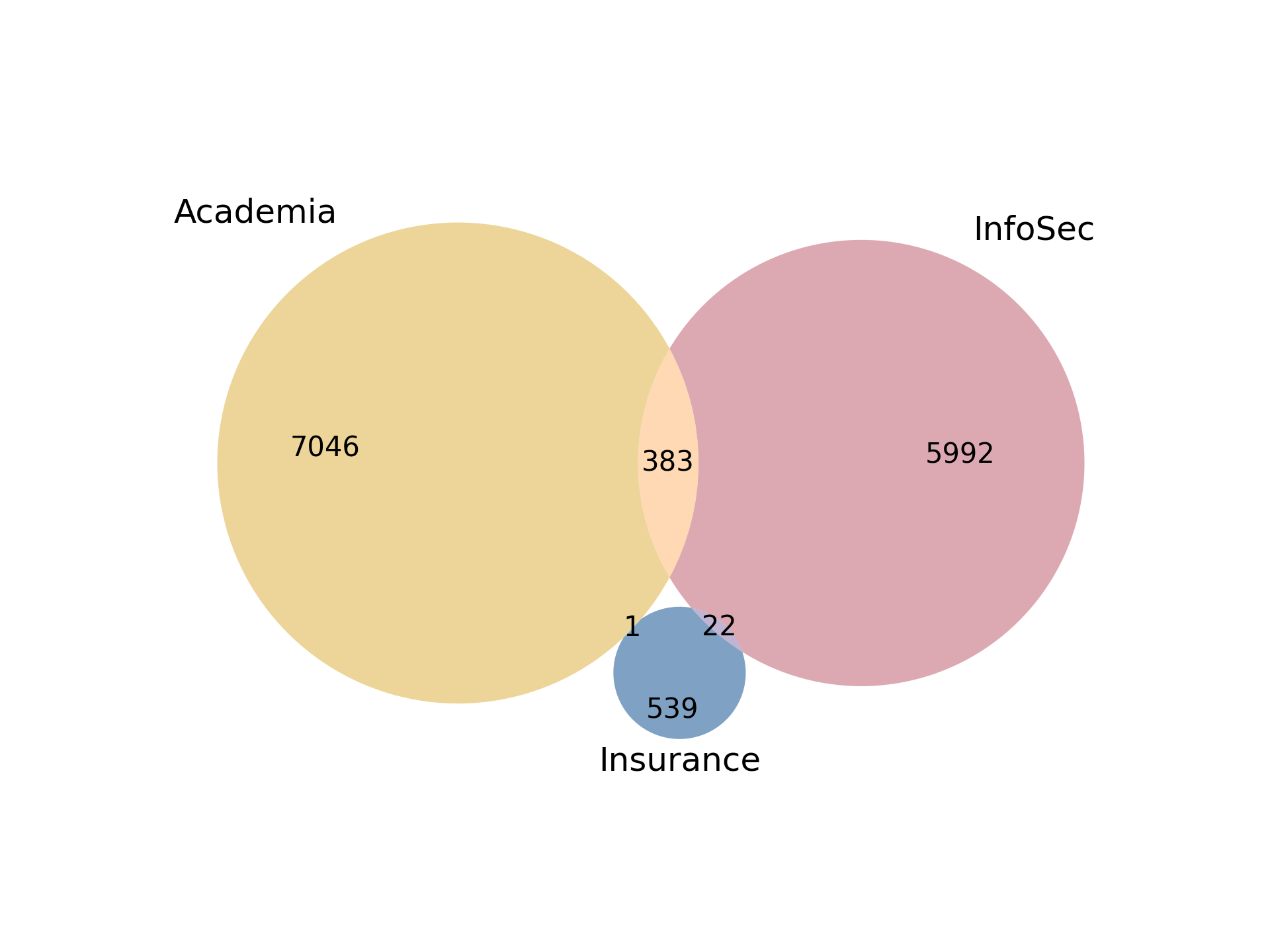}
    \caption{Overlap of speakers across conference types. The Venn diagram shows intersections among industry, InfoSec, and insurance venues during 2014–2022.}
    \label{fig:vennspeakers}
\end{figure}

This subsection  examines the distribution of abstracts among specific individuals.
At industry conferences, the individual typically presents at the conference, sometimes joint with others. 
However, just one author typically presents at academic conferences.
We consider each type of talk separately to account for this, using ``author'' as the terms for individuals associated with academic talks.
Throughout we assume that each unique name corresponds to a single unique person. 
Figure~\ref{fig:vennspeakers} shows there is minimal overlap of speakers between industry, academic, and insurance conferences. 
For this reason, we proceed by analyzing each type of conference separately.

Figure~\ref{fig:freqanalysis} displays the individuals with the most associations with talks, weighted by $\frac{1}{n}$ where $n$ is the number of authors on a paper or individuals on a panel.
Patrick Wardle is a dramatic outlier with 33 talks including more than ten talks at each of Black Hat and DEFCON. 
The rest of the top 30 industry information security speakers gave between 10 and 15 talks across all four conferences. 
Some industry speakers presented exclusively at one conference, such as Patrick Hevesi at Gartner.
No other Top 30 speaker gave a talk at Gartner. 
The top insurance speaker had seven talks in total and just 19 speakers gave more than three talks.

Turning to the academic talks, the top authors have over fifty publications, although rarely as a single author. 
Academics are more likely to publish at multiple conferences than industry speakers. 
For example, all the top 30 authors have presented at IEEE Security and Privacy, USENIX Security and ACM CCS, although some authors did not present at NDSS. 
This kind of analysis raises the question of how speaking slots are distributed among participants.

Table~\ref{tab:conferences_gini} displays the Gini index for each conference.
A score of 0 means all speakers gave the same number of talks, while a score of 1 means one individual gave all talks.
Academic conferences are similarly equal to industry conferences when publications count as $\frac{1}{n}$ where $n$ is the number of co-authors/co-panelists, whereas academic conferences appear considerably less equal when each publications counts as $1$ (see Table~\ref{tab:conferences_gini}).
This corrects for academics who make small contributions to high $n$ papers.
Both metrics agree that NDSS is the most equal academic conference, while ACM CCS is the least.


\begin{figure}
    \centering
    \includegraphics[width=0.5\textwidth]{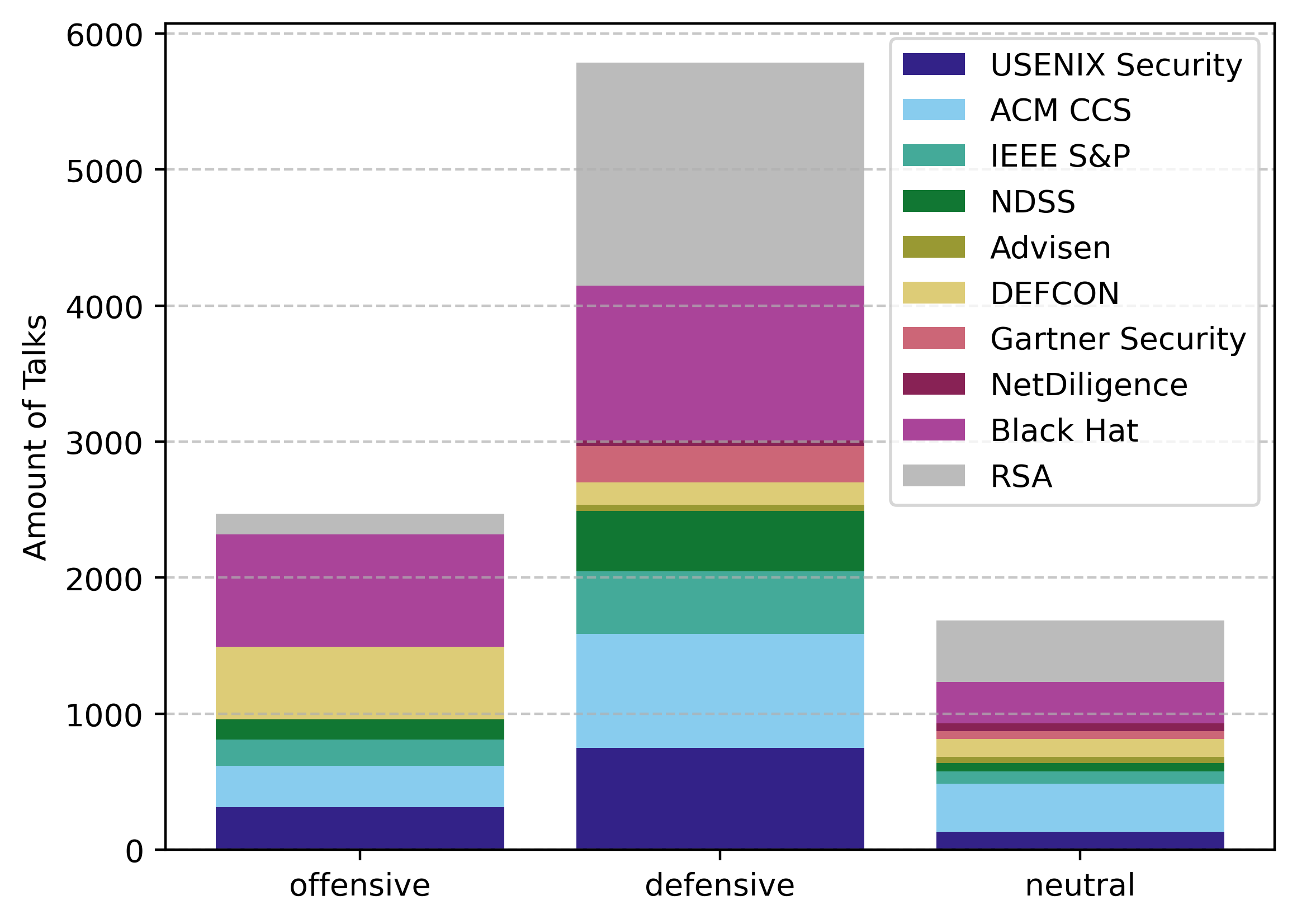}
    \caption{Distribution of offensive, defensive, and neutral talks at each conference from 2014 to 2022.}
    \label{fig:ratio_off_def_neut}
\end{figure}

\subsection{Content of Talks}
\label{subsec:content}
We now explore the topics of abstracts. 
Figure~\ref{fig:ratio_off_def_neut} shows that defensive talks are more common than offensive and neutral talks at all conferences apart from DEFCON, which is explicitly set up for hackers to present their research. At DEFCON, 64\% of the talks were offensive and just 15\% (20\%) of the talks were defensive (neutral). 
Across both cyber insurance conferences, just 1\% of talks were classified as offensive, whereas 55\% of insurance talks were neutral and 44\% were defensive. 
The academic conferences all accepted more defensive work than offensive, and some even accepted more neutral than offensive talks.

\subsubsection{Defensive Talks}
\begin{figure}[t]
    \centering
    \includegraphics[width=0.5\textwidth]{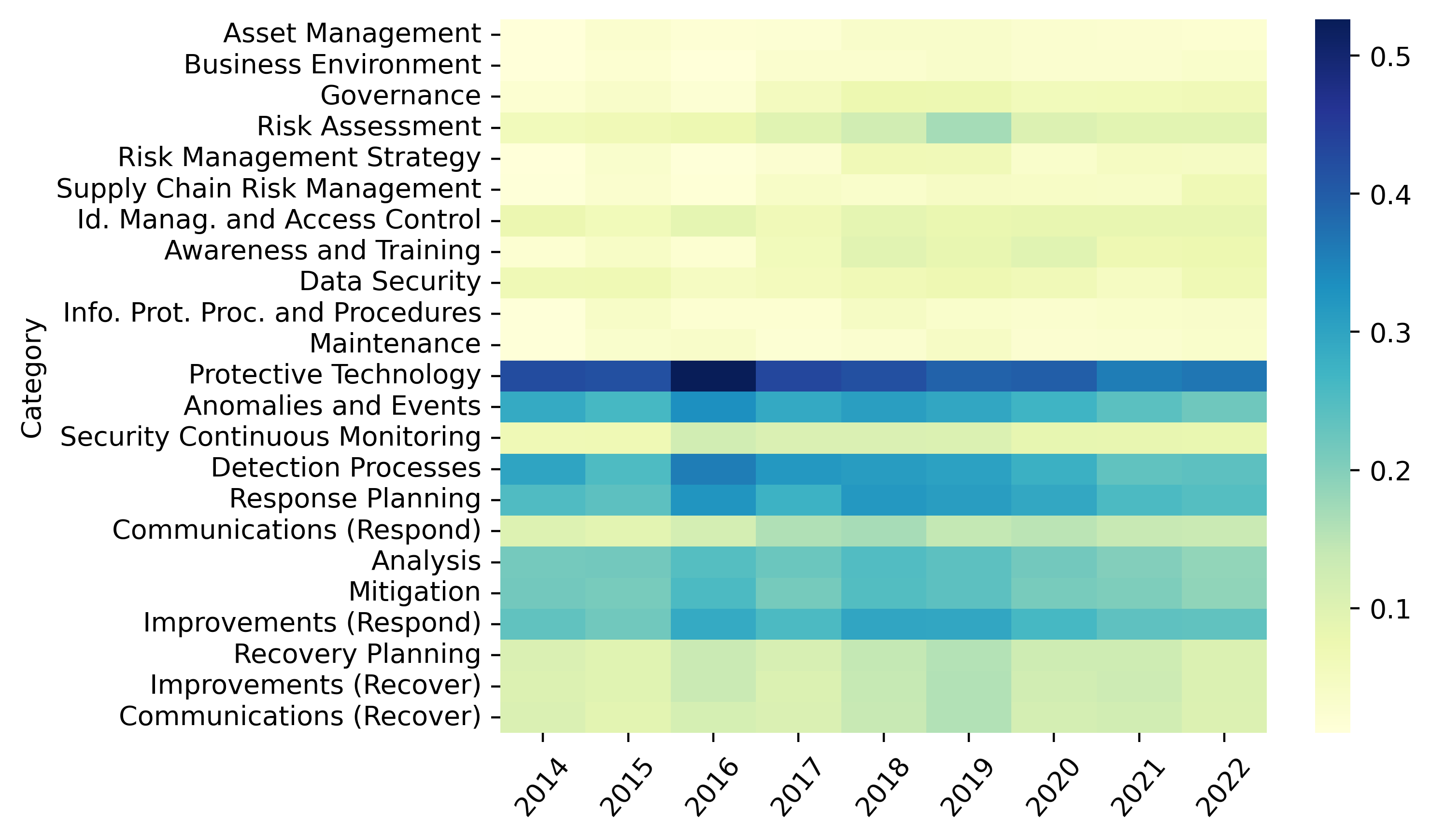}
    \caption{Similarity score of defensive talks mapped to the NIST Cybersecurity Framework over time (2014–2022).}
    \label{fig:NIST_years}
\end{figure}

Figure~\ref{fig:NIST_years} shows that the similarity of talks across areas of NIST CSF has remained relatively stable over time.
There is a slight trend of increasing abstracts about risk assessment and governance, as well as decreasing similarity to protective technology.
This is generally true when considering trends in topics at individual conferences, and we display some such figures in the Appendix.
For this reason, we proceed by analyzing the abstracts aggregated across all years.
The NIST Cybersecurity Framework is divided into five core functions: \textbf{Identify}, \textbf{Protect}, \textbf{Detect}, \textbf{Respond} and \textbf{Recover}.

\begin{figure}[t]
    \centering
    \includegraphics[width=0.5\textwidth]{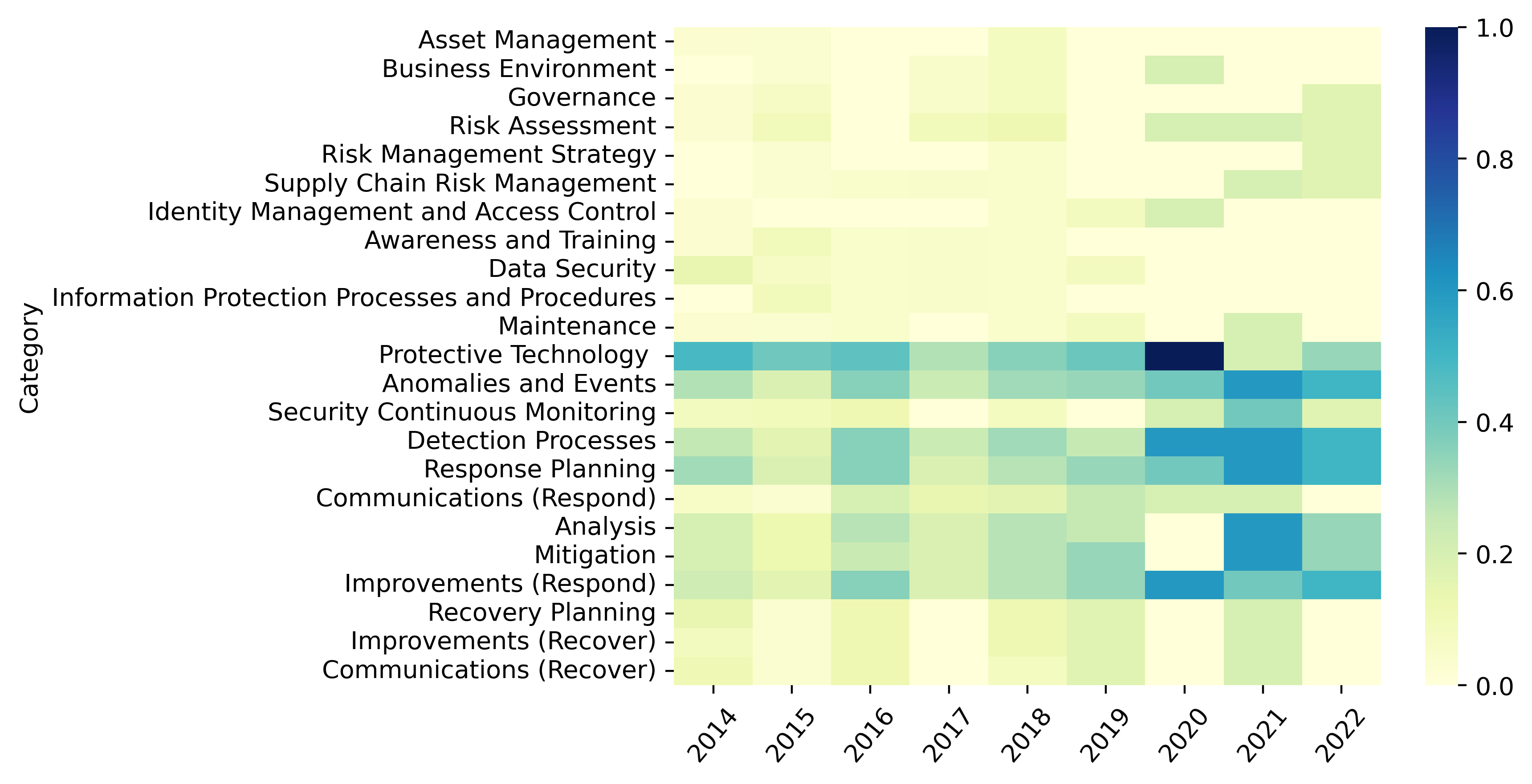}
    \caption{Similarity score of DEFCON defensive talks mapped to the NIST Cybersecurity Framework over time (2014–2022).}
    \label{fig:NIST_years_defcon}
\end{figure}

Before analyzing the aggregate data, there is one exception to the general lack of a temporal trend in topics. Figure~\ref{fig:NIST_years_defcon} shows how DEFCON displayed a considerable temporal shift towards Detection and Response functions of NIST CSF.
This could be evidence of the conference's commercialization, which is an ongoing topic of discussion.
The equivalent figures for other conferences can be found in the Appendix.

\begin{figure}[t]
    \centering
    \includegraphics[width=0.5\textwidth]{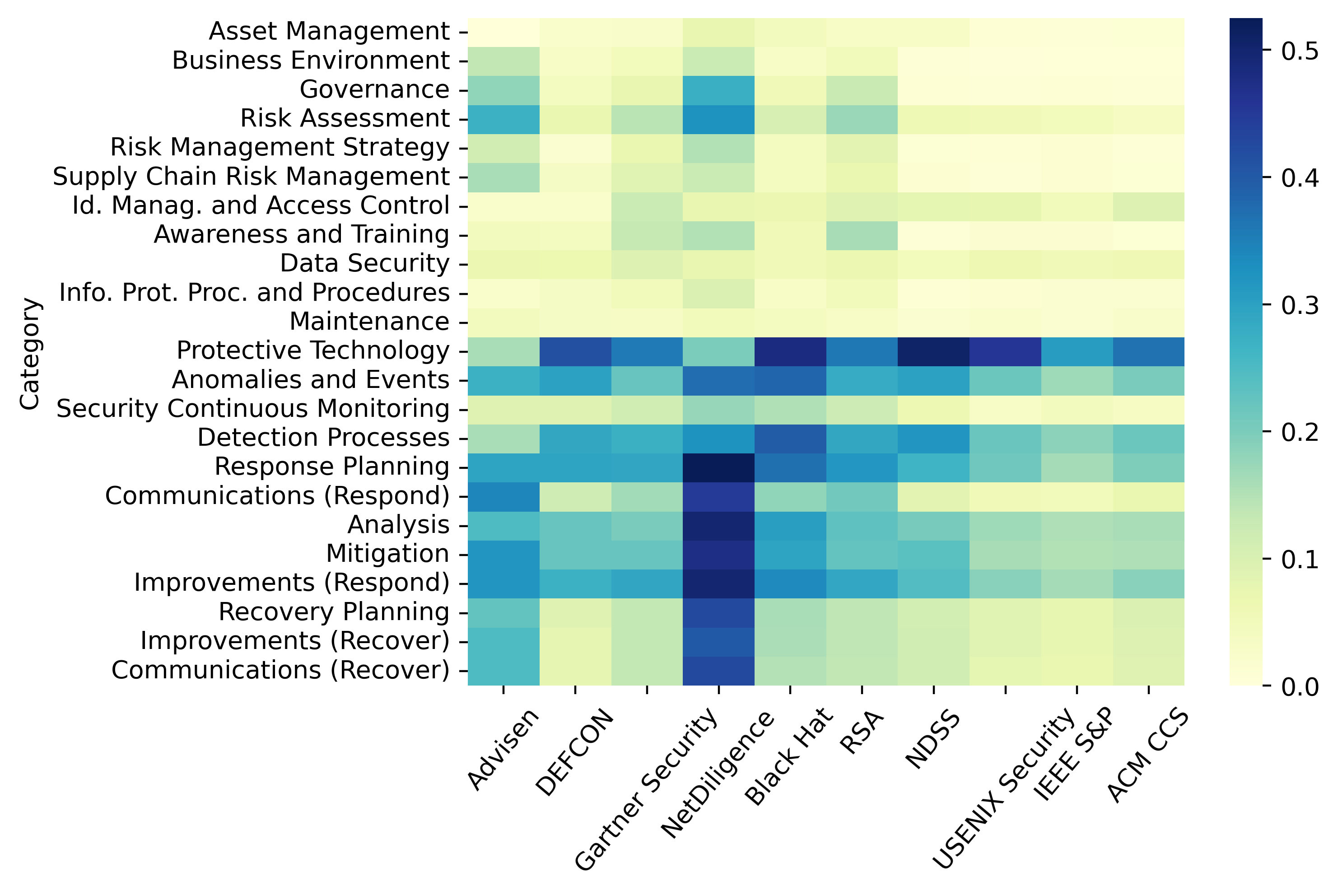}
    \caption{
    Similarity score of defensive talks mapped to the NIST Cybersecurity Framework over time (2014–2022).}
    \label{fig:NIST-cosine}
\end{figure}

\paragraph{\textbf{Identify}}
The identify function comprises the first six rows in Figure~\ref{fig:NIST-cosine}. Broadly speaking, \emph{Identify} covers how organizations make security decisions. Both industry InfoSec and insurance talks address these topics more than the academic talks.
The insurance conferences and Gartner display high similarity to \emph{Risk Assessment}, which captures the likelihood and impact of different security incidents, and \emph{Governance}, which cover the process of establishing a security policy. It is also notable that the research-oriented industry information security conferences (DEFCON and Black Hat) display less similarity to \emph{Identify}.
Notably, there is low similarity to the final two sections:
\emph{Risk Management Strategy} involves understanding the firm's meta-reasoning like the assumptions and risk tolerance of the firm, meanwhile \emph{Supply Chain Risk Management} involves assessing the risk of external firms.
Academics study this area when they conduct threat modeling~\cite{kaur2025threat, usman2025sok}.

Researchers should not dismiss the \emph{Identify} function as `non-technical' and outside the scope of computer security. Many of the \emph{Identify} tasks can be solved using technical data.
For example, a recent SoK on asset discovery highlights how \emph{Asset Management} can be approached using network measurements~\cite{vermeer2021sok}. Similarly, \emph{Risk Assessment} is analogous to cyber incident prediction, which has been studied as a machine learning problem based on technical indicators and published at top-tier venues~\cite{liu2015cloudy, bilge2017riskteller}.

\begin{figure}
    \centering
    \includegraphics[width=0.4\textwidth]{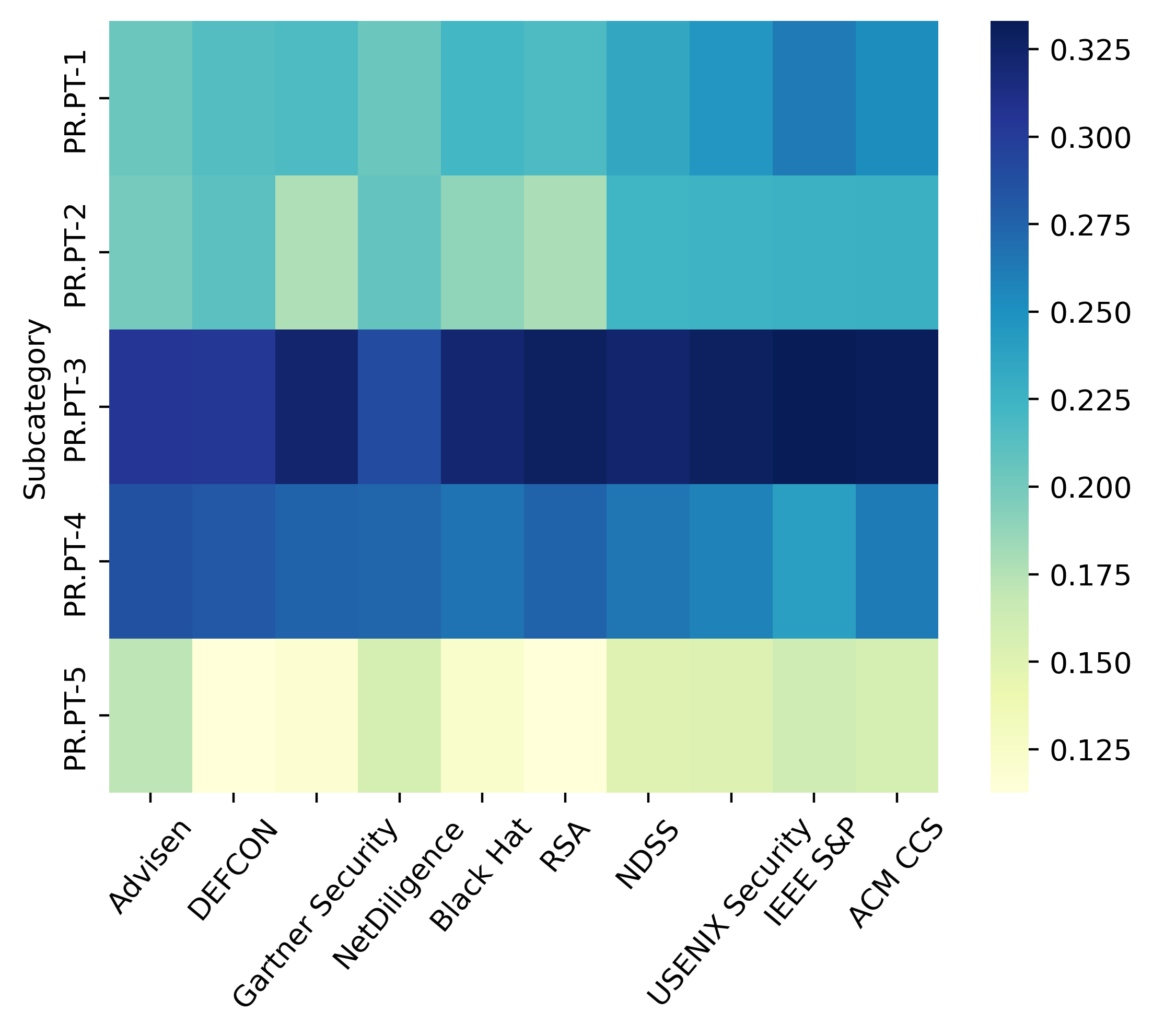}
    \caption{Cosine similarity score for talks within the \emph{Protective Technology} function of the NIST Cybersecurity Framework from 2014 to 2022.}
    \label{fig:NIST-pro-tech}
\end{figure}

\paragraph{\textbf{Protect}}
The protect function covers the next six rows in Figure~\ref{fig:NIST-cosine}. Broadly speaking, \emph{Protect} describes technical controls and organizational processes that directly prevent incidents from occurring. 
Of these six categories, all conferences had the most similarity to \emph{Protective Technology}.
Figure~\ref{fig:NIST-pro-tech} displays cosine similarity to each sub-category of \emph{Protective Technology}. The two sub-categories with the highest similarity concern system hardening (PR.PT-3) and network security (PR.PT-4). The next two subcategories concern audits and logging (PR.PT-1), and removable media (PR.PT-2). Finally, resilience mechanisms like fail-safe and load balancing (PR.PT-5) display the lowest similarity.

Academic talks display some similarity to \emph{Identity Management and Access Control}.
This include the evergreen research topic that is passwords~\cite{morris1979password, bonneau2012quest}. 
The \emph{Awareness and Training} category received more coverage in the industry talks relative to the academic talks~\cite{ho2025understanding}. Academic research related to this may instead be presented at conferences focusing on Human-Computer Interaction like SOUPS and Chi~\cite{garfinkel2014usable}. 

The remaining three categories had low similarity. 
\emph{Maintenance} had the lowest similarity across all of the defensive categories. 
This finding supports the observation that researchers often overlook maintenance even though it is crucial to successful operations~\cite{collier2020cybercrime}. 
The \emph{Information Protection Processes and Procedures} category consists of procedures to ensure availability like back-ups, recovery plans and so on.
Industry talks more frequently cover these topics. 
Low similarity to \emph{Data Security} was perplexing given this is so core to computer security. 
The first two sub-categories of \emph{Data Security}, namely data at rest and in transit, are core academic topics.
However, the other sub-categories are organizational procedures that tend not to be covered at academic conferences.

\paragraph{\textbf{Detect}}
The next three rows concern network monitoring and incident detection. 
Broadly speaking, all of the industry conferences displayed moderate similarity to \emph{Security Continuous Monitoring}, \emph{Anomalies and Events}, and \emph{Detection Processes}. These three categories respectively correspond to: what monitoring technology is in place; how anomalies are identified and responded to; and the surrounding organizational processes like testing, communication, and improvement. 
NDSS displays higher similarity than the other academic conferences to all three \emph{Detect} categories, which likely results from the conference's focus on network security. 
Across the five NIST Cybersecurity framework functions, the abstracts were most consistently similar to \emph{Detect}.

\paragraph{\textbf{Respond}}
The respond function comprises the next five rows, which are broadly the steps taken to investigate and contain an incident. This does not capture coordinated response, such as botnet take downs~\cite{vu2025assessing}. 
Abstracts were most similar to \emph{Response Planning}, which involves establishing incident response procedures before the incident occurs. The talks also displayed reasonable similarity to \emph{Mitigation}, which involves actions to contain and resolve an incident, and also \emph{Improvements}, which is about extracting lessons learned from incidents. This is perhaps surprising given the consensus that the security community struggles to extract lessons from incidents~\cite{knake2021learning}.

There was low similarity with the  \emph{Analysis} category, which includes forensic investigations and vulnerability disclosure policies, especially at academic conferences. This could be because researchers present at specialist venues like \emph{Digital Investigation}~\cite{garfinkel2010digital}. 
There was relatively low similarity to \emph{Communications}, which involves coordinating with internal and external stakeholders like law enforcement, especially at academic conferences. 

Much like with \emph{Detect}, NDSS displays higher similarity to than the other academic conferences. The insurance conferences, especially NetDilligence, display remarkably high similarity to all five categories. 
The high level of similarity holds even for \emph{Communications} despite the other conferences largely ignoring this topic, which supports the theme that insurers have a different perspective on security.

\paragraph{\textbf{Recover}}
The final function comprises three categories that all happen post-incident, which all have equivalent categories in the \emph{Response} function.
All conferences show less similarity to the \emph{Recover} category.
This suggests conference talks are more focused on short-term tasks like containing the incident than the long-term goal of restoring the victim to previous function, which is the goal of \emph{Recover}.
The Advisen conference displays the most similarity, which makes sense given recovery is the goal of insurance.

\subsubsection{Offensive Talks}
We mapped offensive talks to the MITRE ATT\&CK framework. This consists of 14 tactics and each tactic has multiple techniques. 
Gartner, Advisen and NetDilligence are not displayed due to a lack of offensive talks.

\begin{figure}[t]
    \centering
    \includegraphics[width=0.5\textwidth]{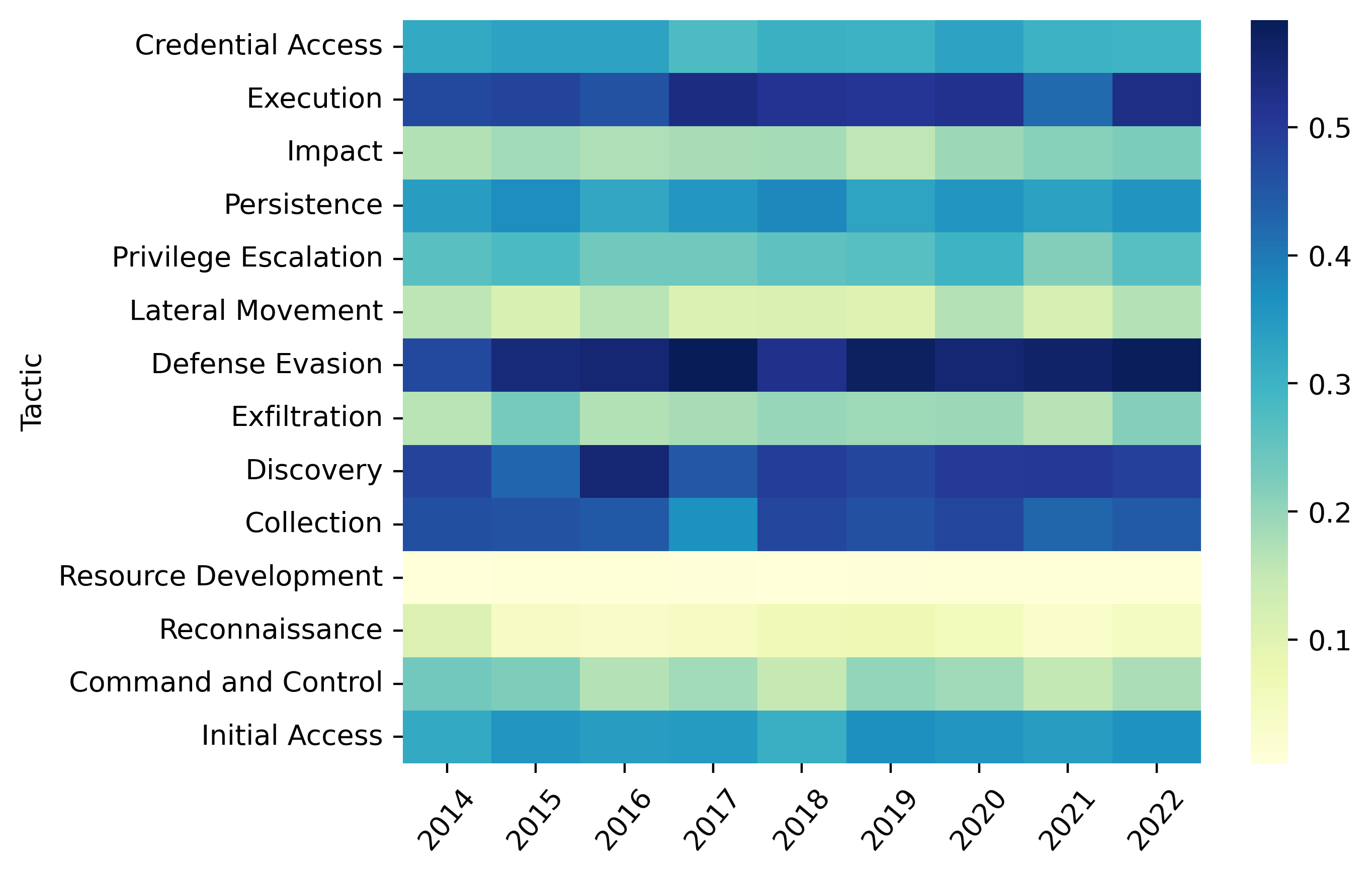}
    \caption{Similarity score of offensive talks mapped to the MITRE ATT\&CK framework over time (2014–2022).}
    \label{fig:MITRE_years}
\end{figure}

Much like with defensive talks, Figure~\ref{fig:MITRE_years} displays no clear temporal trend in which offensive tactics are covered by conference abstracts. There is, however, a subtle trend towards \emph{Impact}. 
We group the 14 tactics into temporal categories: \textbf{Pre-compromise}, \textbf{First Compromise}, \textbf{Expansion}, and \textbf{Outcome}. 

\paragraph{\textbf{Pre-Compromise}}
The \emph{Reconnaissance} and \emph{Resource Development} tactics involve collecting information and other resources (e.g.\,acquiring exploits) before the attack is launched.  \emph{Resource development} is the tactic that the conferences are least similar to. \emph{Reconnaissance}, which involves techniques like active scanning and gathering information on the victim's network, is also very rarely covered by talks.
These tactics may be infrequently discussed because they are non-technical.

\begin{figure}
    \centering
    \includegraphics[width=0.5\textwidth]{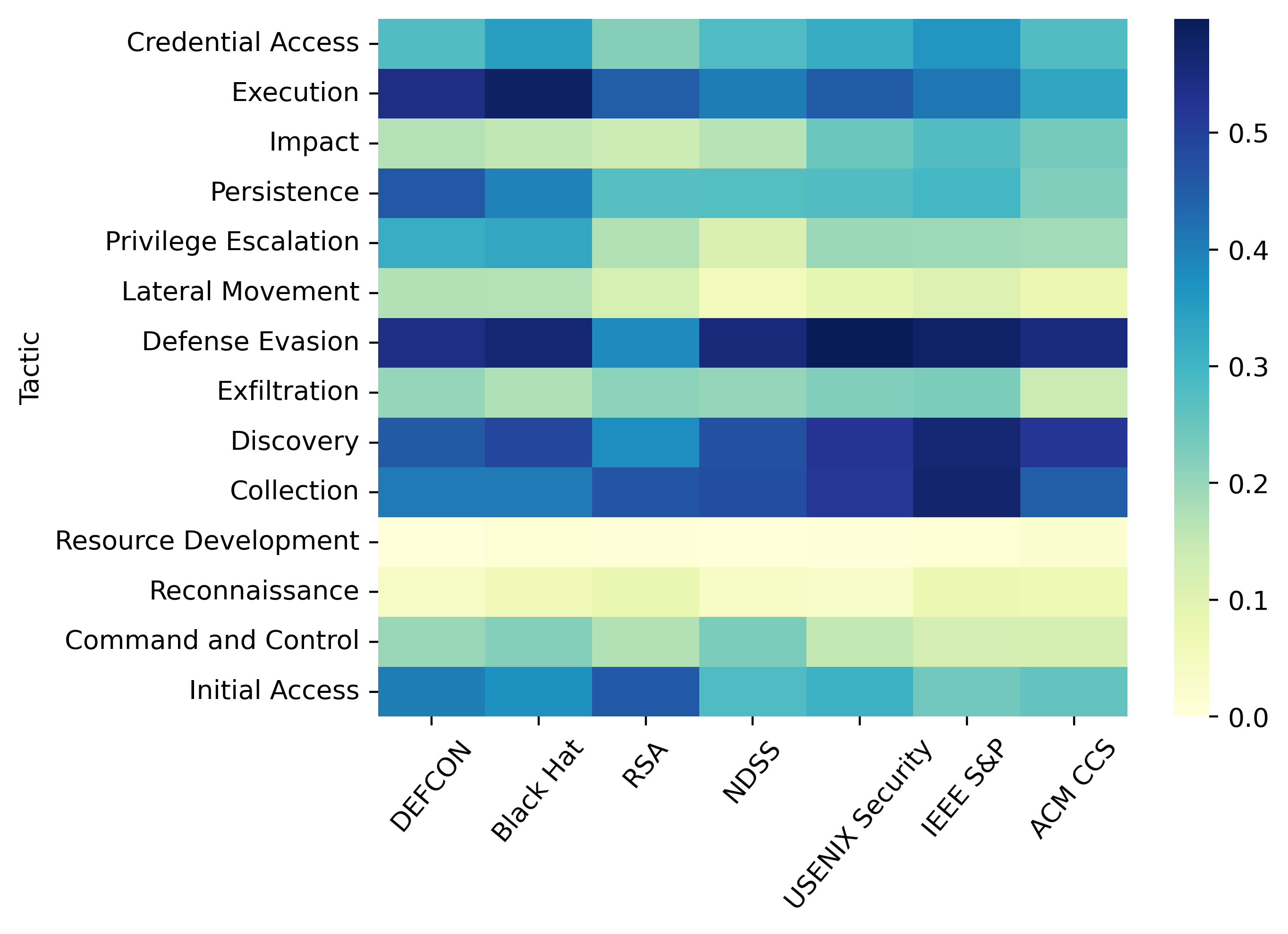}
    \caption{Similarity score of offensive talks mapped to the MITRE ATT\&CK framework across conferences.}
    \label{fig:MITRE}
\end{figure}

\paragraph{\textbf{First Compromise}}
The next four tactics concern how the attacker compromises the first system on a network. \emph{Initial Access} describes the entry vector into the network, which includes different types of phishing and account compromises. The industry talks conferences display moderate similarity to \emph{Initial Access}. USENIX Security abstracts are most similar to this among the academic conferences. This breaks from the trend that NDSS talks were most similar to the NIST Cybersecurity Framework out of the academic conferences.

\begin{figure}
    \centering
    \includegraphics[width=0.5\textwidth]{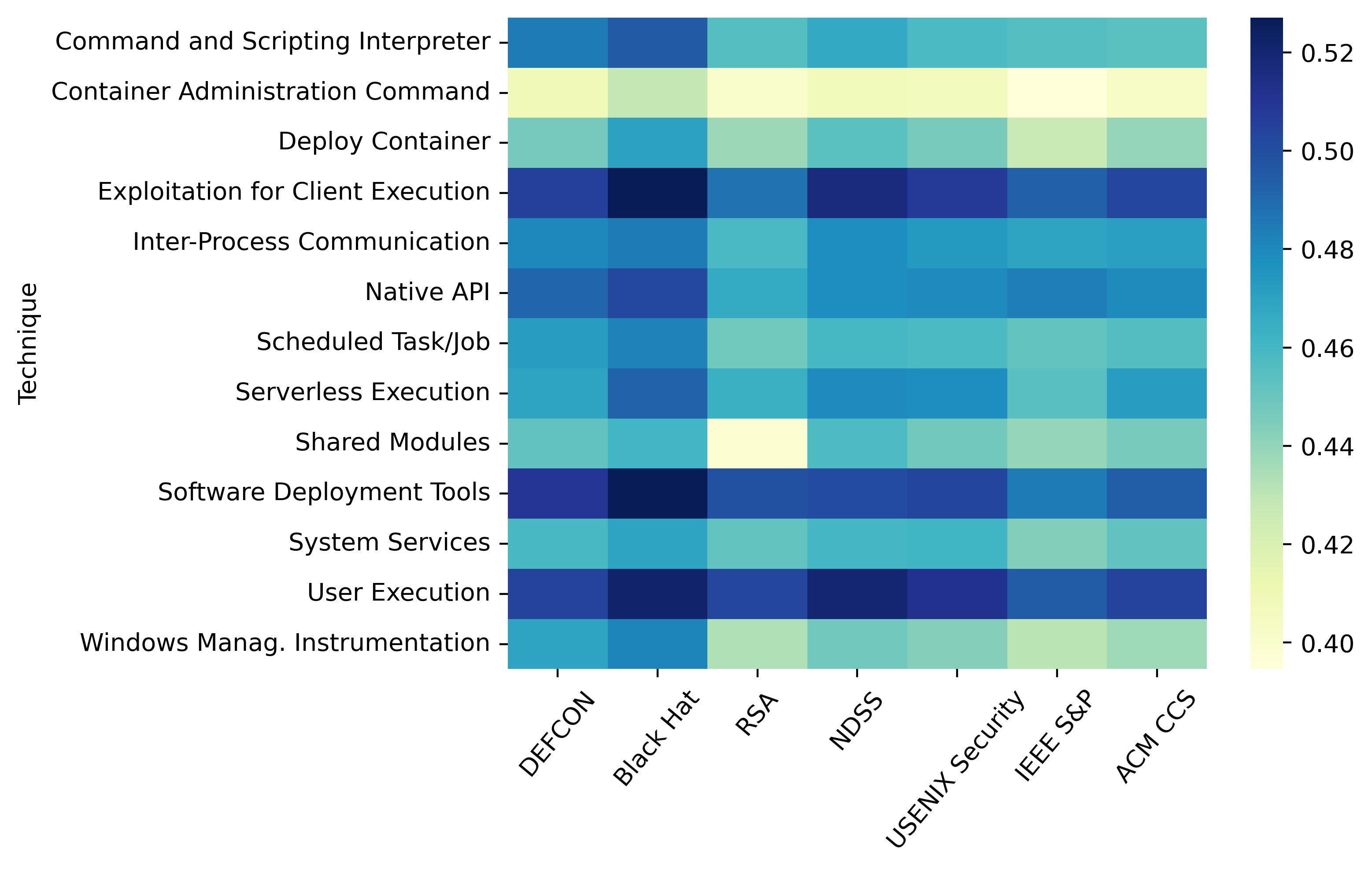}
    \caption{Cosine similarity score for talks classified under the \emph{Execution} technique category of MITRE ATT\&CK framework across conferences.}
    \label{fig:MITRE-execution}
\end{figure}

\emph{Execution} describes tactics that attackers use to run malicious code on the victim's systems.
The industry talks display the most similarity to this category, in particular the Black Hat talks. The academic talks are highly similar to this category, but it is not the highest. Figure~\ref{fig:MITRE-execution} displays the similarity of the tactics within \emph{Execution}.

\emph{Persistence} describes how attackers maintain access across events like the user rebooting, switching user or any other interruption. It is discussed most often at DEFCON, with Black Hat talks the second most similar. The other conferences display relatively low similarity. This suggests that academic researchers are most concerned by the initial vulnerability, and not how the adversary maintains access to the system.

\emph{Privilege Escalation} techniques allow an attacker to gain more permissions to access files and other system resources. This technique shows a similar pattern to \emph{Persistence}, in which academic talks display low similarity relative to industry conferences. 
USENIX Security talks have the most similarity out of the academic conferences.

\paragraph{\textbf{Expansion}}
The next three steps concern how the attacker moves through a victim's network. \emph{Defense Evasion} concerns how the adversary avoids detection. The talks display a high similarity to this and also \emph{Discovery}, which concerns how the adversary collects information about the victim's network. Unlike the previous tactics, the academic talks display higher similarity to both than the industry talks.

\emph{Credential Access} involves stealing account names and passwords, such as via keylogging. The conference talks are only moderately similar to this, with academic talks showing more similarity than industry.
The conference talks display low similarity to the third tactic in this category, namely \emph{Lateral Movement}. It covers how the adversary moves through the victim's network.

\paragraph{\textbf{Outcome}}
These tactics concern what was achieved by the compromise, although these techniques could also be used to expand access. 
\emph{Collection} concerns identifying and gathering data on the victim network.
It is the tactic in this category with the highest similarity to the conference talks.
Meanwhile, \emph{Exfiltration} concerns how the data is stolen from the network, which has reasonably low similarity to the conference talks.

The \emph{Command and Control} tactic involves establishing communication channels to maintain control over the infected systems. This has moderate similarities across both industry and academia.
The NDSS conference displays the highest similarity, again because this is fundamentally a problem of distributed computing, albeit an illegal one. 
Finally, the \emph{Impact} technique concerns an effort to ``manipulate, interrupt, or destroy your systems and data'', which would involve ransomware for example. The academic talks display higher similarity to this category, although it is moderate-high for all conferences apart from RSA.

\section{Threats to Validity and Limitations}
\label{sec:limitations}
Our study has several limitations that affect the interpretation and generalizability of the results.

\textbf{Threats to External Validity (Generalizability)} \emph{Missing Data:} Our comparative longitudinal study was impacted by changes in reporting and missing data. 
We could not collect some years of talks due to data unavailability.

Our longitudinal dataset is affected by occasional missing years or incomplete sponsor information, particularly for some industry conferences. For example, the RSA conference was created in 1991 but we could only extract the program from 2017 onward. Similarly, there was a lot of missing sponsorship information, which led to a long list of exceptions to the sponsors column in Table~\ref{tab:conferences_sponsorship}. These omissions mean that 13 of the 74 expected conference-year sponsorship data points (approximately 17.6\%) are missing from the analysis. This lack of data systematically biases aggregate statistics toward the better-documented conferences and limits the completeness of the longitudinal trend analysis. Future releases of our dataset will incorporate missing years as additional data become publicly available.

\textbf{Threats to Construct Validity (Measurement)} \emph{Real Names:} Counting authors can be complex because two unique names may correspond to the same individual (e.g.\,Emma Jones and Emma Q. Jones), and one name may correspond to multiple individuals (e.g.\,Emma Jones from New York and Emma Jones from California). Additionally, some authors may use pseudonyms rather than their real names. 
For simplicity, we counted unique names as unique people. 
We considered using affiliations to distinguish between two unique individuals with the same name, but researchers switch institutions too often for this to be reliable.

Similar issues are present for company names. Company names are inconsistently abbreviated (Hewlett Packard vs HP), subsidiaries and even departments are listed (e.g.\,both Microsoft Corporation and Microsoft Research were listed as sponsors), and some companies change names (e.g.\,Facebook to Meta). We fixed such issues manually where we could, but we likely missed instances for less well known companies. This fragmentation is estimated to affect less than 2\% of total unique entities after manual correction, primarily impacting measures of speaker concentration and sponsor overlap but not materially changing overall long-term trends.

\emph{Classification:} Mapping abstracts to categories from security frameworks is challenging because: (i) the task is inherently subjective; (ii) the categories are high-level with fuzzy boundaries; (iii) abstracts typically focus on specific technological systems; and (iv) even state-of-the-art LLMs are known to have errors, let alone the version we used that is now far from state-of-the-art. These factors help to explain why the agreement between the two NLP techniques (see Section~\ref{subsec:validation}) was disappointing. Ultimately, we focused on the LLM-based mapping because it was the best available technique at the time of the initial analysis and it captures how  language is used in real discussions in the LLM's training set. This was also confirmed by the results of manual annotation: 264 abstracts (2.7\% of the corpus) were validated, achieving a 92.4\% agreement (95\% CI ±3.1 pp) and a Cohen's kappa coefficient of 0.89 with human annotators. However, residual errors and category ambiguities are still possible. This can be contrasted against the cosine similarity approach, which is trained on a bag-of-words extracted from official documents.
However, it is not unreasonable to prefer a different mapping, such as using another NLP technique or another cybersecurity framework. To enable this, we open-source our data and analysis scripts~\cite{walter_2024_15989593}.

\emph{Conference Coverage:} Our comparative analysis (\textbf{RQ3}) would have been more interesting if we collected data from more conferences and also went further back in time. For example, the FIRST Conference on Computer Security Incident Handling would likely have covered more topics related to \emph{Respond}, but from a more technical perspective than the insurance conferences. The core challenge to adding more data is that we could only semi-automated our scrapers in order to ensure data quality, and had to create a new parser for each conference. This made it costly to add additional conferences and also to scrape further back in time. Again, we hope that open-sourcing our data and analysis will allow future work to extend our work by adding new conferences.

\textbf{Threats to Internal Validity (Causality)}:
Our results are descriptive, not causal. While we discuss possible explanations for observed patterns (e.g., topic prevalence or speaker concentration), these should not be interpreted as evidence of causation. Since our study is based on descriptive data, we explicitly avoid making causal claims to prevent threats to internal validity.

\section{Discussion}    
\label{sec:discussion}
This section discusses the results and how our framing assumptions influenced the analysis.

\textbf{Speakers} Academic conferences display greater information sharing across venues than industry. Many industry speakers present at only one conference, and speakers rarely cross the insurance--InfoSec line (see Figure~\ref{fig:vennspeakers}).
It is unclear whether industry conferences would benefit from more cross-pollination.
The call for broader information sharing is appealing, however it could be that more efficient communication is enabled by the shared language and assumptions within each community.

A common question is whether sponsorship buys influence~\cite{drezner2017ideas}.
Conferences with more sponsors have fewer words per abstract, which could be interpreted as sponsored conferences involving less complex talks. 
However, the relationship between sponsorship and speaker selection appears limited. For example, at Black Hat — the event with the largest number of sponsors — 63\% of sponsors never had an affiliated talk, and 85\% of speakers’ organizations never sponsored the conference.
Nevertheless, sponsorship patterns raise concerns for academic conferences. Despite focusing on privacy, two of the five most frequent sponsors of academic conferences directly profit from behavioral advertising. The relationship between BigTech companies and researchers has been called into question by other researchers~\cite{abdalla2021grey}.

A similar question is whether it is a problem that a small number of speakers/authors hold the majority of influence.
For illustrative purposes, we can compare the Gini index for conferences (see Table~\ref{tab:conferences_gini}) to the index of income inequality in specific countries.
The distribution of talks across speakers is typically around 0.37--0.44, which is close to the post-tax distribution of income in relatively unequal societies like the US (0.375) and Mexico (0.42).
All prestigious industry and academic talks have higher inequality than Germany (0.296).\footnote{All figures sourced from the OECD using post-tax income: \url{https://stats.oecd.org} (accessed 19 October 2025)}
Inequality in academic talks is not necessarily a cause for concern. Scientific conferences should provide a platform for the `best' science, and that likely means the `best' scientists will present more often. However, an unequal distribution can also result from a narrow perspective of what constitutes `high-quality' research. 
Among academic venues, ACM CCS displayed the highest inequality across both Gini computation approaches.

\textbf{Topics} Expanding the topical diversity of academic security conferences could mitigate inequality by allowing new researchers to present alternative perspectives. Some areas - particularly the \emph{Identify} function within the NIST Cybersecurity Framework - remain underrepresented.
Understanding current security management practices~\cite{hielscher2023employees} could inspire technical approaches to management problems. For instance, \emph{Asset Management} can be inferred from network traffic~\cite{dekoven2019measuring}; \emph{Risk Assessment} can be framed as ML prediction~\cite{liu2015cloudy}; and \emph{Governance} can involve technical audits~\cite{rahaman2019security, mahmud2020cardpliance}.

Turning to offensive talks, their distribution is much more even across the techniques in MITRE ATT\&CK, spanning both academic and industry conferences. The one exception is information gathering that occurs pre-compromise, which is rarely discussed at conferences. Preventing resource development is perhaps better conducted by governments, such as by disrupting criminal markets for exploits~\cite{allodi2015then}.
The lack of offensive talks at insurance conferences raises the question of whether there is a sufficient understanding of attacker operations.


\textbf{Variance}
Insurance-focused events are clear outliers: they emphasize management and resilience rather than technical controls or attacker techniques. This confirms the finding that insurers are more focused on post-incident response than pre-incident mitigation~\cite{wolff2022cyberinsurance}.
Within MITRE ATT\&CK mappings, the only offensive conference (DEFCON) is comparatively focused on how compromise was achieved (execution, persistence and initial access), while the academic conferences focus on activities that take place after compromise. 

Over time, there appears to be a gradual shift towards the \emph{Identify} function and away from \emph{Protective Technology} and \emph{Detect} functions.
This trend contradicts the prediction of Schneier that the 2010s would be the decade of response~\cite{schneier2014future}. Instead, there appears to be a subtle shift to security management. However, the apparent declines should not be overstated, as the `declining' categories remain the most frequently discussed in absolute terms.

\textbf{Assumptions}
In mapping abstracts to cybersecurity frameworks, we imply that there should be balanced coverage across framework categories.
Our core assumption is that a broader range of speakers and topics implies a healthier research ecosystem.
While diversity has known benefits, specialization also drives depth and innovation. This tension reflects the classic explore–exploit trade-off, where the objective — maximizing global security knowledge — remains undefined.
We acknowledge that reasonable readers may disagree with our emphasis on explore.

More fundamentally, it is unclear whether encouraging specific lines of work is appropriate, especially given the sources of those frameworks.
This has the potential to impinge on academic freedom.
Although we explained why ATT\&CK and NIST CSF were the best among the alternatives (see Section~\ref{sec:methodology}), it is entirely unclear whether they are `good enough' in terms of granularity, completeness and so on.
This may be actively contested by some.

\textbf{Future Work} We focused on speakers but did not consider listeners/readers.
Doing so is crucial to understanding the impact of conference presentations.
Collecting this information risks introducing some form of surveillance that undermines the privacy of attendees. 
One privacy-preserving approach is to scrape the number of views that uploaded video recordings received and to study the comments. 
Alternatively, an (imperfect) academic proxy for impact on readers is citation count, but the industry equivalent requires more sophisticated research designs.

To address the limitations described in  Section~\ref{sec:limitations}, future work should add extra conferences and map abstract to different security frameworks and compare the results.
Further, a qualitative study collecting data from conference organizers could explore hidden dynamics, such as negotiations with industry sponsors.
This would provide insights into unseen details like what sponsors demand from academic conferences.
Gaining access to industry conference organizers is considerably more difficult.

\section{Conclusion}    
\label{sec:conclusion}
Security conferences are an important but understudied channel by which knowledge and information is shared. Our paper built a longitudinal dataset of talks at 10 prestigious conferences across industry and academia. We characterized who speaks at these conferences (\textbf{RQ1}), about which topics (\textbf{RQ2}), and how this compares across conferences and over time (\textbf{RQ3}).
To answer the latter two research questions, we mapped conference abstracts to cybersecurity frameworks using two NLP techniques, which show moderate agreement.

The results show academics moved freely between the big four venues, but just 
5\% of academic authors also presented at InfoSec or Insurance conferences. 
The equivalent figure was 4\% for insurance speakers, and 7\% for InfoSec speakers.
This suggests there is limited information sharing across academia, InfoSec and insurance; a potential area for improvement.
We also showed there is an unequal distribution of talks across speakers, with this inequality varying across conferences.
In terms of sponsors,  academic conference are dominated by BigTech firms, whereas InfoSec conference sponsors are predominantly cybersecurity vendors.
Unequal distribution of speaking time or sponsorship sources can be a sign that parties have the power to influence which security topics are studied and which solutions are adopted.

Defensive talks were more common than offensive and neutral ones at all conferences, except for DEFCON, where offensive topics dominated due to its hacker-focused nature.
This allows us to quantify the balance of offensive and defensive papers at academic conferences, which have been criticized for focusing on attack papers~\cite{herley2017sok}.
Offensive abstracts are distributed evenly across the categories of MITRE ATT\&CK, apart from \emph{Reconnaissance} and \emph{Resource Development}. 
Defensive abstracts focused on \emph{Protective Technology} and categories related to monitoring.
Academic conferences are less likely to focus on the \emph{Identify} and \emph{Recover} functions from the NIST Cybersecurity Framework. 

These findings support a more nuanced understanding of security information sharing, in which success should be evaluated with regards to specific channels and institutions. Researchers have been disappointed by the quality of information shared via online sources~\cite{acar2016you, fischer2017stack} and threat intelligence feeds~\cite{li2019reading, bouwman2020different, bouwman2025can}.
However, there is more success when it comes to vulnerability management. Bug bounties incentivize the discovery of new information~\cite{finifter2013empirical, sridhar2021hacking}. Further, information about fixes is widely available, the problem lies in acting on the information by installing patches~\cite{frei2006large, li2017large}. Our results suggest security conferences should be considered a success story, closer to patch management than threat intelligence sharing. This is perhaps not surprising given attendees, sponsors and research founders invest considerable resources in attending and sponsoring these conferences.
An interesting follow-on study would speak to conference organizers to enrich our quantitative analysis with qualitative insights about the unobservable aspects of conference organization.

Going forward, organizers of prestigious conferences should ensure there is a range of speakers and topics.
To increase diversity of speakers, program committees could explore proposals like: (i) reserving slots for authors/institutions who have not presented before; and (ii) introducing a maximum number of submissions per individual (implemented by USENIX Security, for example).
To increase topic diversity, conferences could edit the call for papers to address under-studied topics according to our results, and possibly lower the bar for acceptance for these topics---a policy not unlike special issues for journals.
We only raise these policies as provocations, acknowledging such policies will be argued to be in tension with the principle of meritocracy.
Nevertheless, we hope our article can prompt discussions among program committees.

\bibliographystyle{unsrt}
\bibliography{references.bib}

\clearpage
\appendix
\section*{Appendix: Supplementary Figures and Tables}

\begin{figure}[h]
    \centering
    \includegraphics[width=0.5\textwidth]{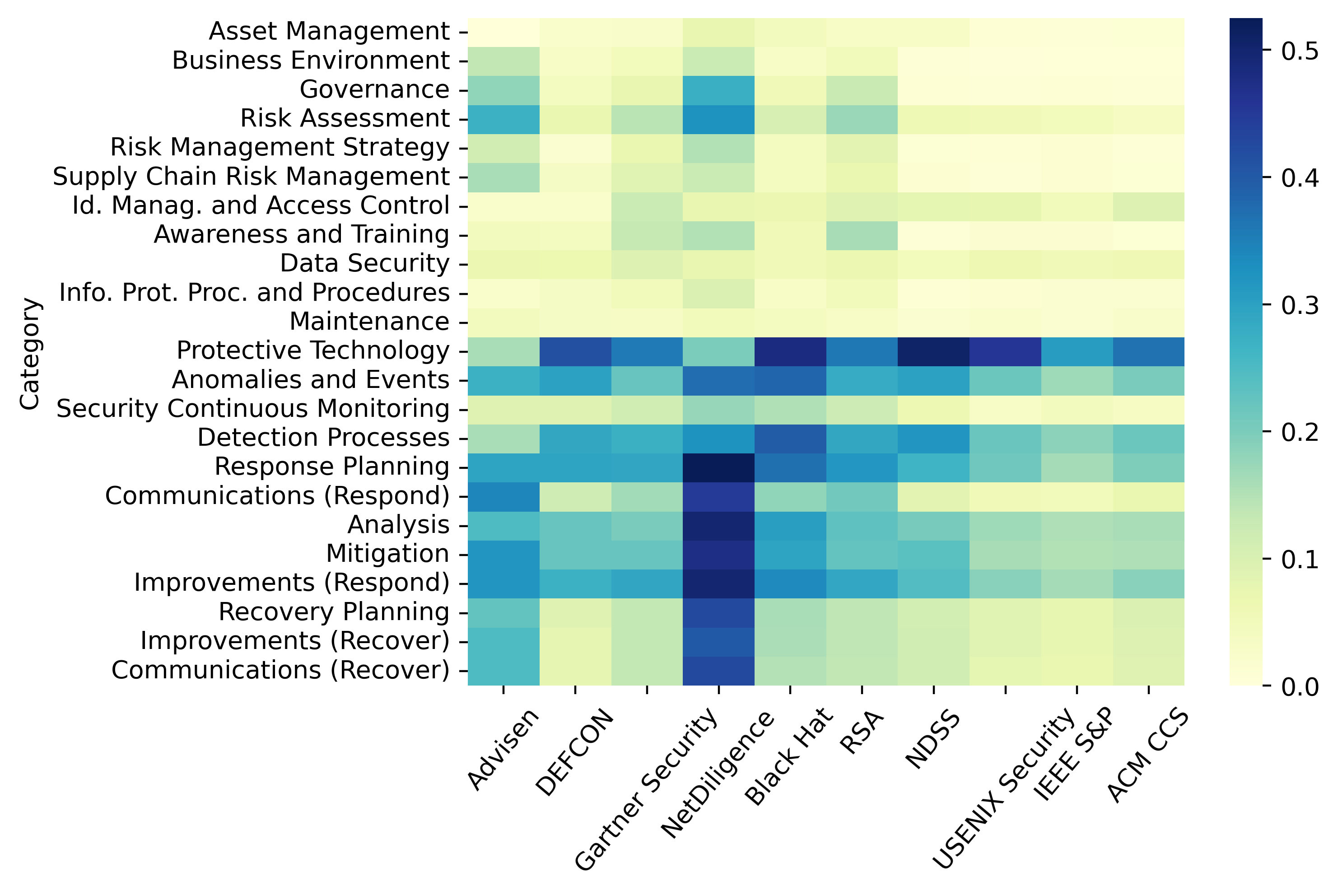}
    \caption{Mean similarity score (averaged across GPT-based and cosine similarity measures) for defensive talks mapped to the NIST Cybersecurity Framework.}
    \label{fig:NIST-cosine-ensemble}
\end{figure}
\begin{figure}
    \centering
    \includegraphics[width=0.5\textwidth]{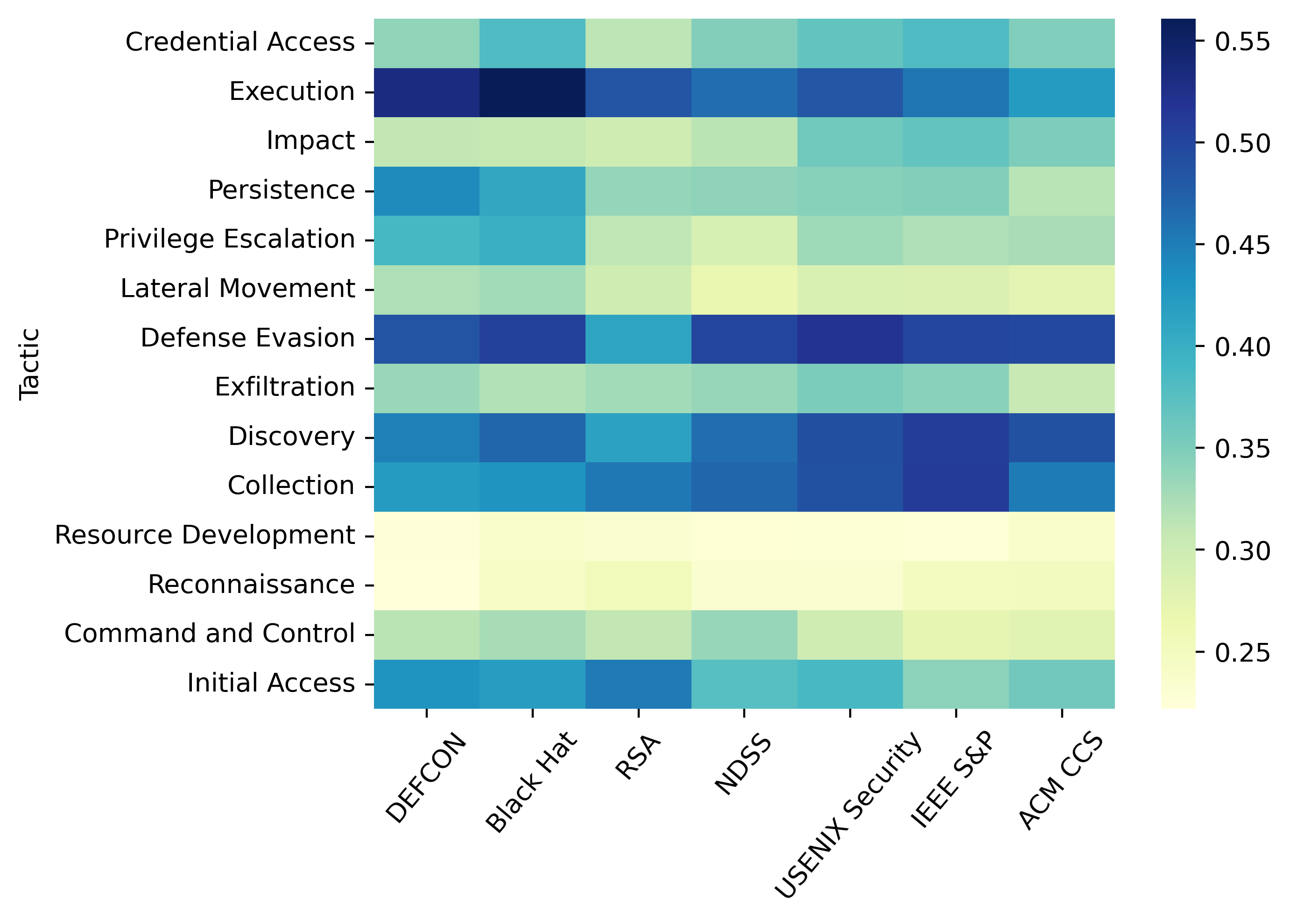}
    \caption{Mean similarity score (averaged across GPT-based and cosine similarity measures) for offensive talks mapped to the MITRE ATT\&CK Framework.}
    \label{fig:MITRE-ensemble}
\end{figure}

\begin{figure}[t]
    \centering
    \includegraphics[width=0.5\textwidth]{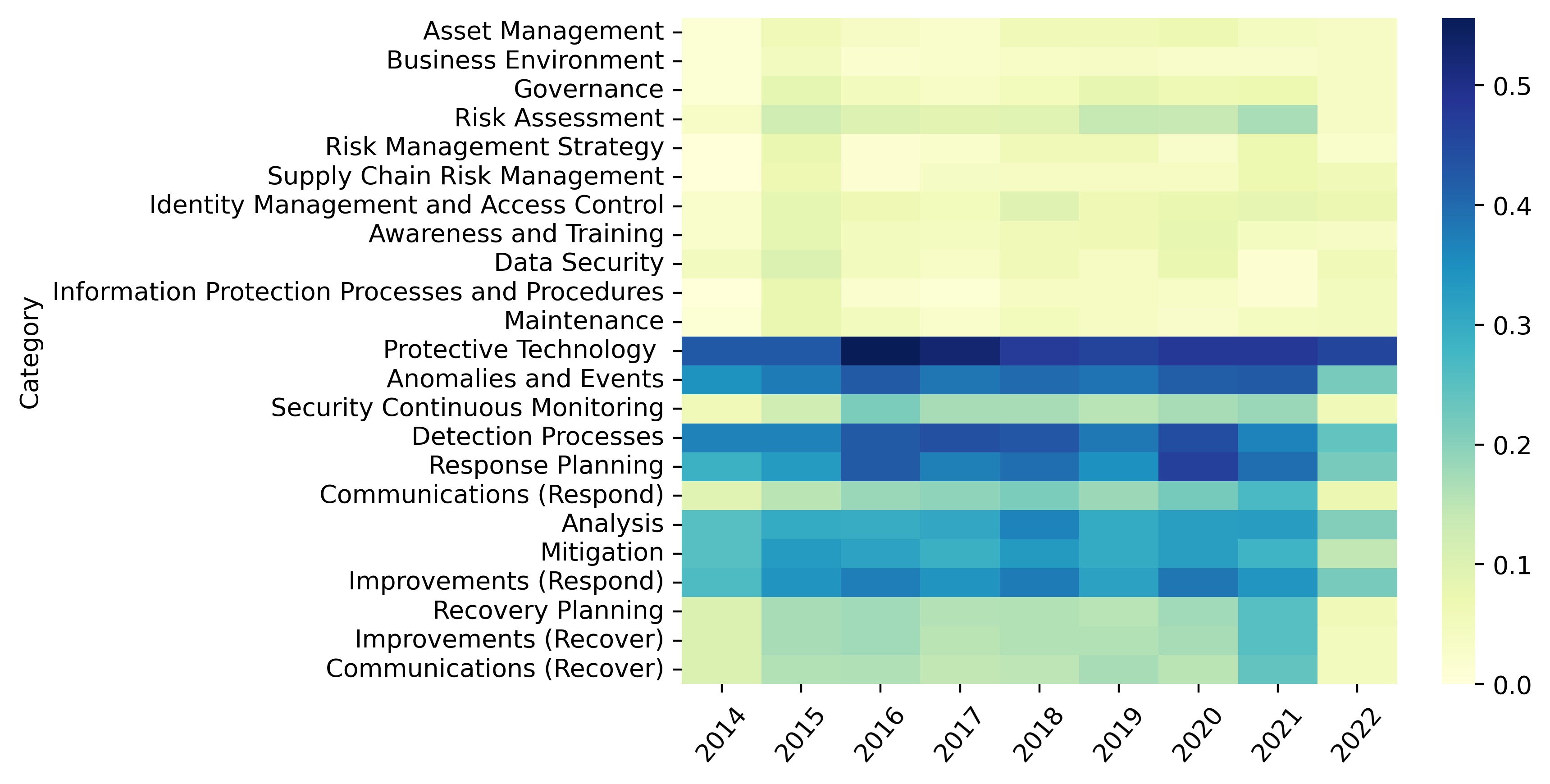}
    \caption{Similarity score of Black Hat defensive talks mapped to the NIST Cybersecurity Framework from 2014 to 2022.}
    \label{fig:NIST_years_blackhat}
\end{figure}

\begin{figure}[t]
    \centering
    \includegraphics[width=0.5\textwidth]{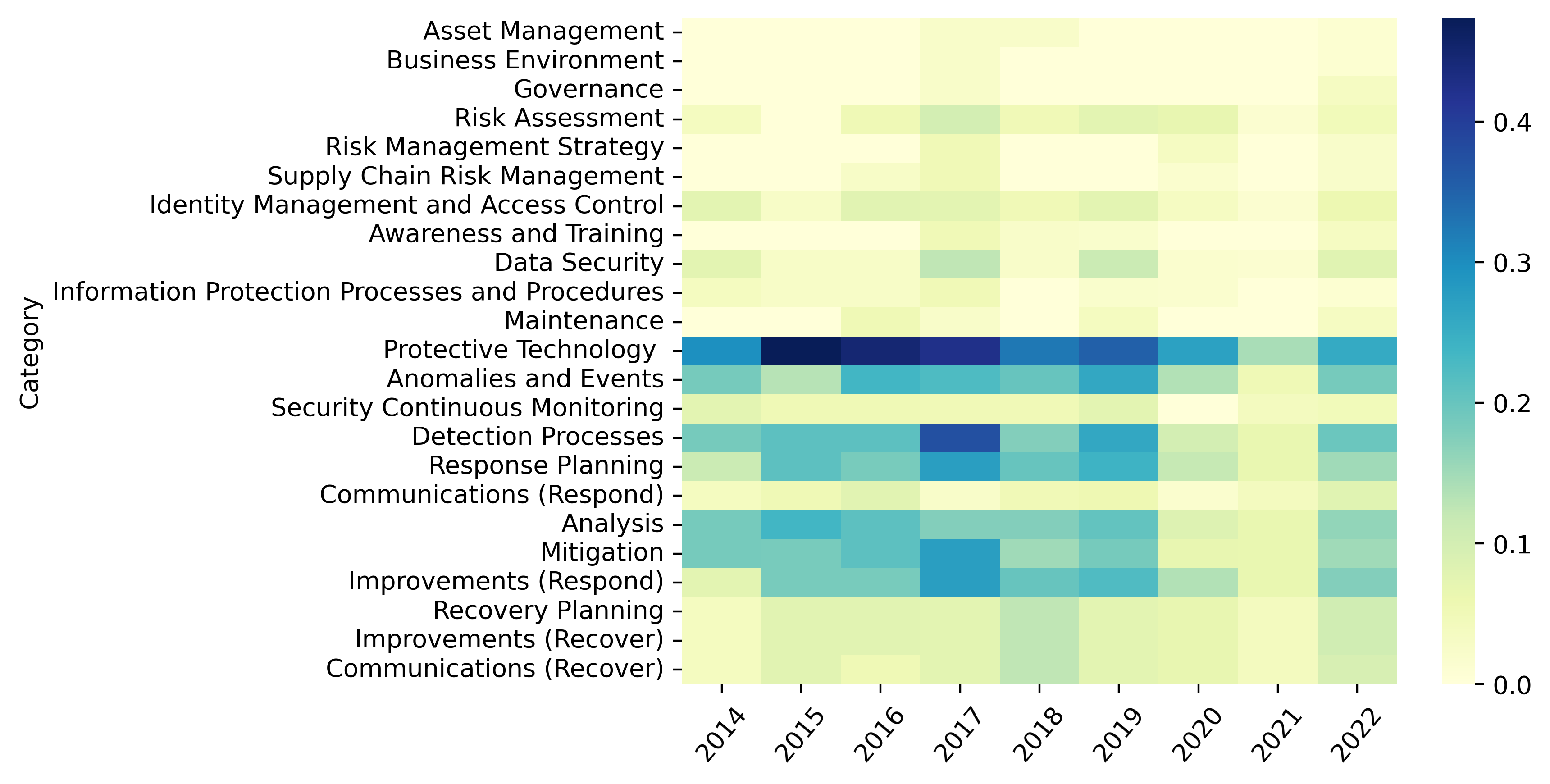}
    \caption{Similarity score of IEEE S\&P defensive talks mapped to the NIST Cybersecurity Framework from 2014 to 2022.}
    \label{fig:NIST_years_ieee}
\end{figure}

\begin{figure*}{}
    \centering
    \includegraphics[width=0.88\textwidth]{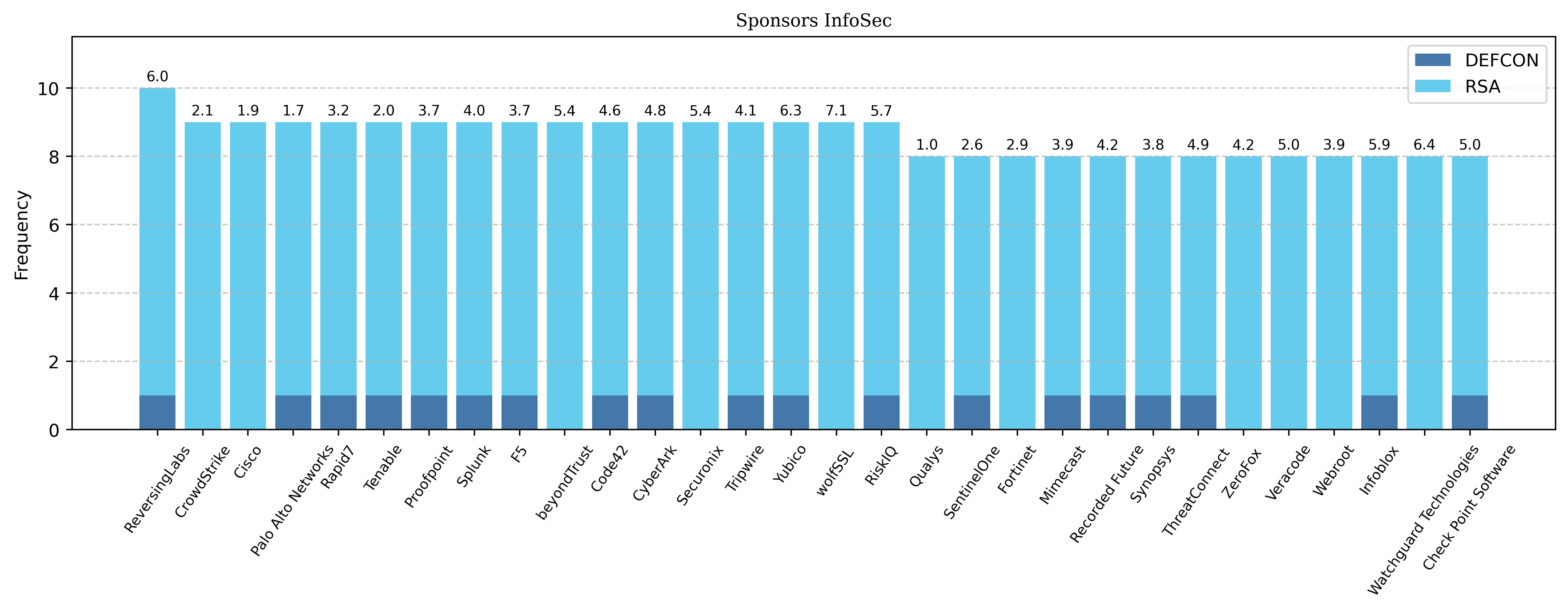}
    \includegraphics[width=0.88\textwidth]{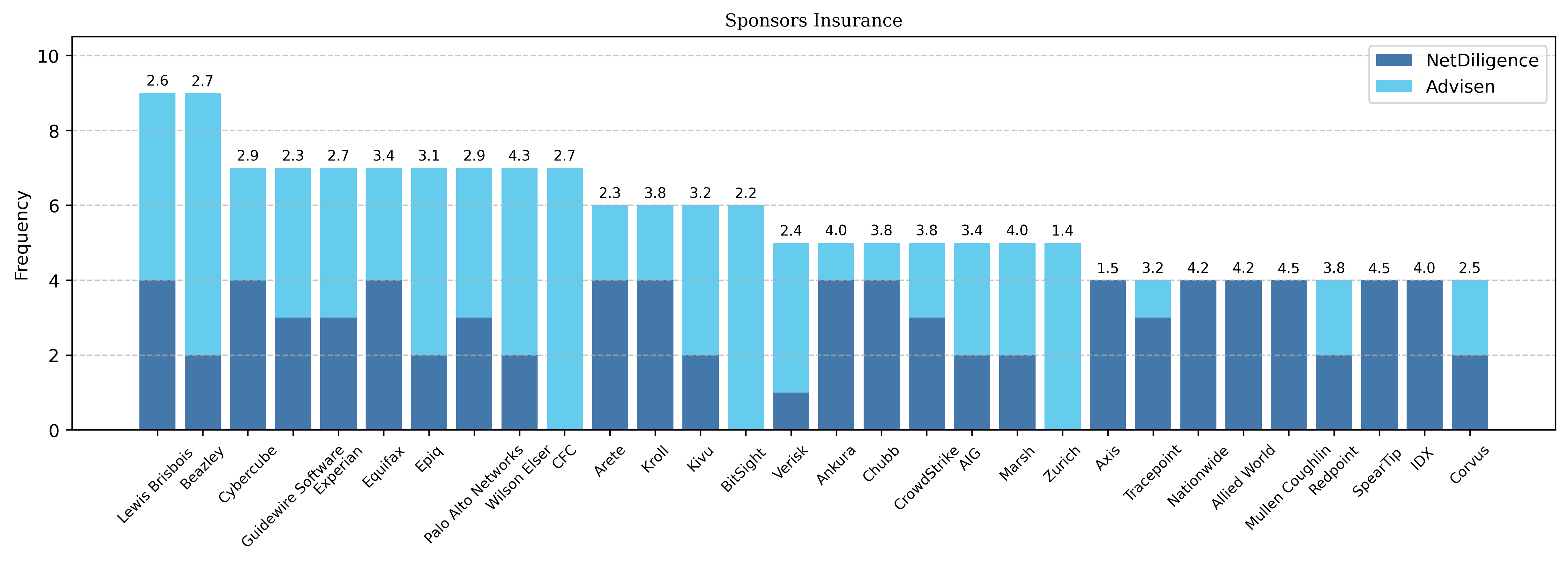}
    \caption{Top sponsors of InfoSec and cyber insurance conferences aggregated over 2014–2022. Numbers above the bars indicate each sponsor’s mean sponsorship tier. Note: sponsor data for some InfoSec conferences were incomplete.} 
    \label{fig:industrysponsorsfreq}
\end{figure*}

\begin{figure}[t]
    \centering
    \includegraphics[width=0.5\textwidth]{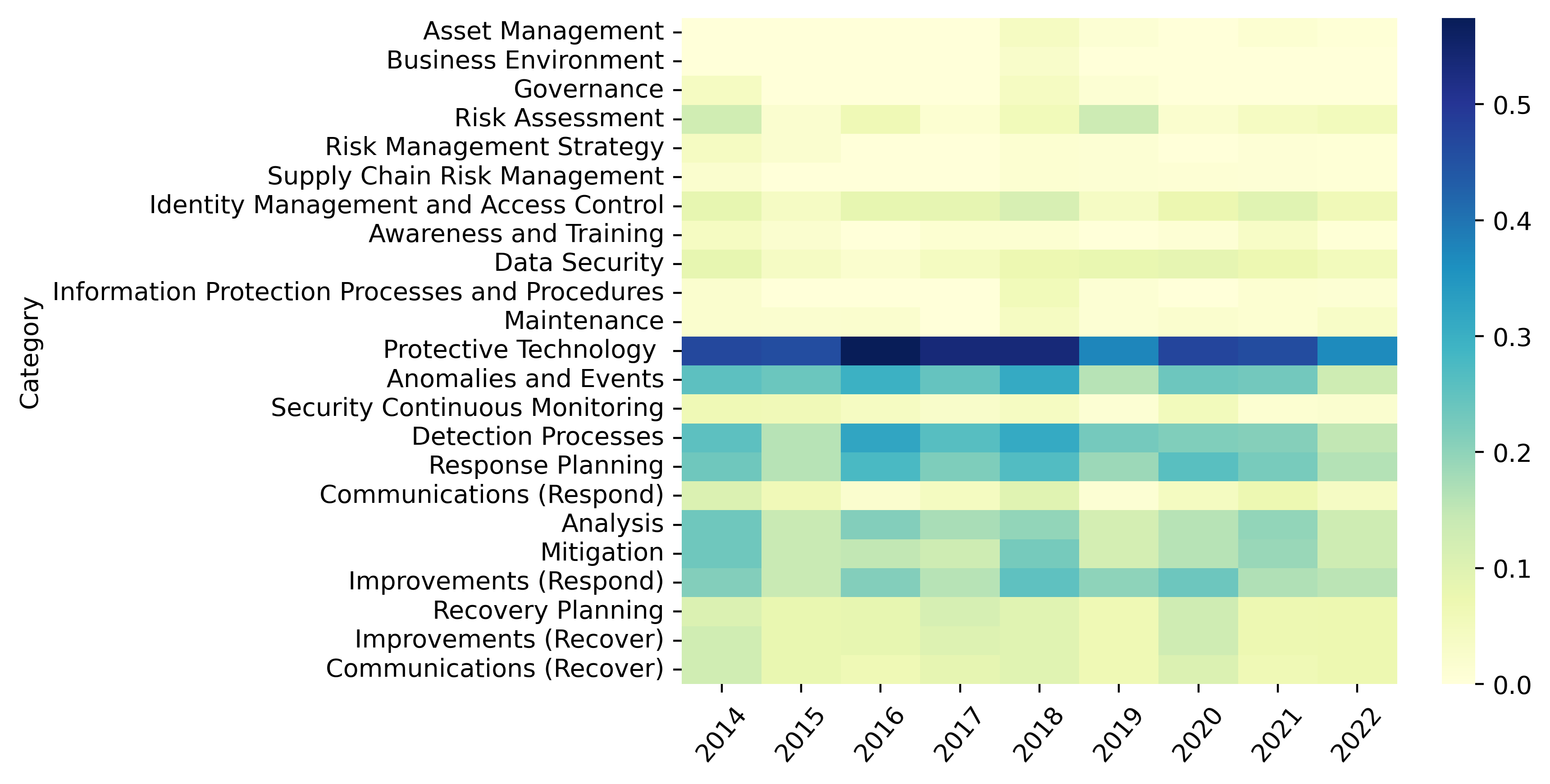}
    \caption{Similarity score of USENIX Security defensive talks mapped to the NIST Cybersecurity Framework from 2014 to 2022.}
    \label{fig:NIST_years_usenix}
\end{figure}

\begin{figure}[t]
    \centering
    \includegraphics[width=0.5\textwidth]{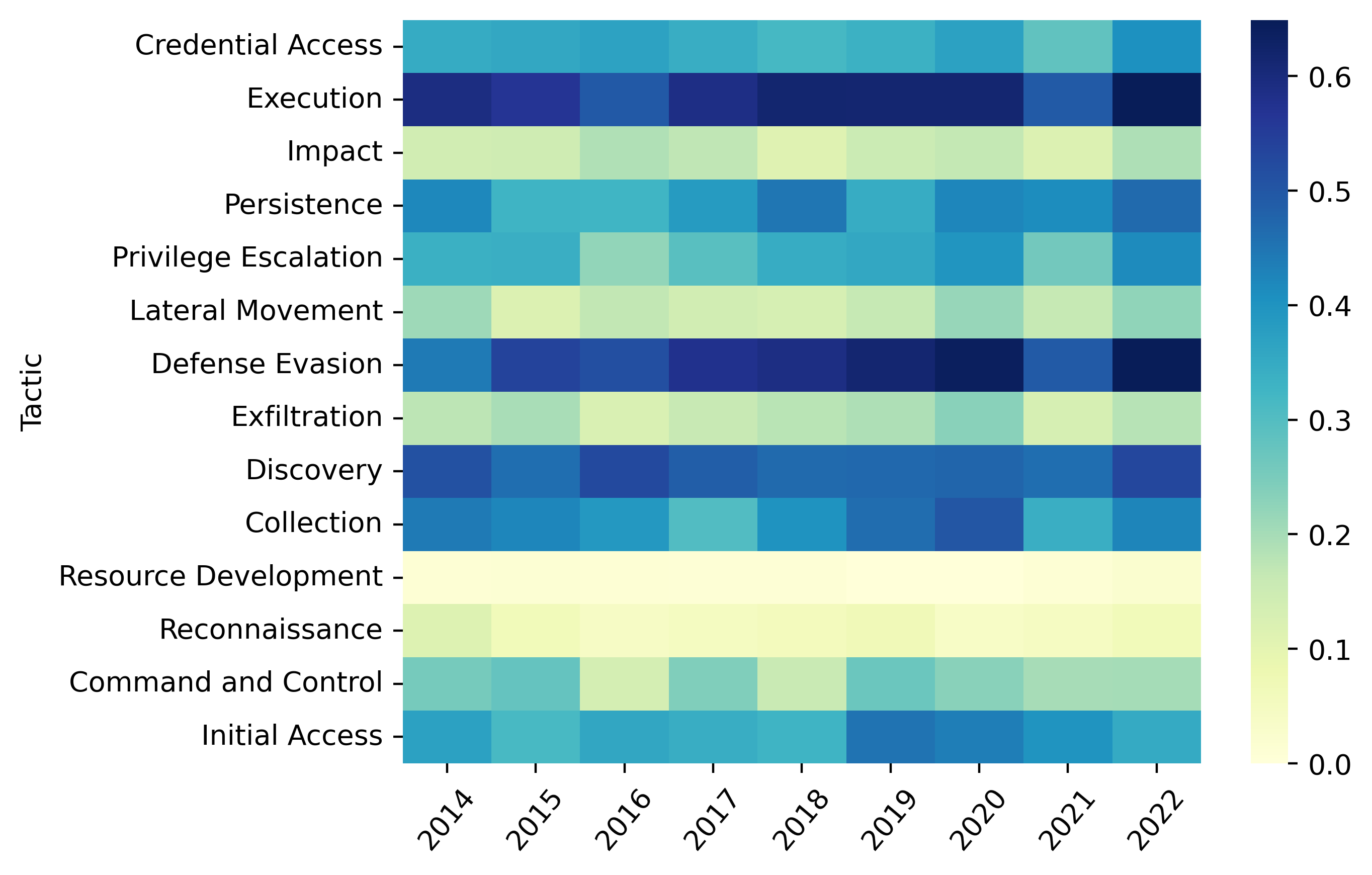}
    \caption{Similarity score of Black Hat offensive talks mapped to the MITRE ATT\&CK Framework from 2014 to 2022.}
    \label{fig:MITRE_years_blackhat}
\end{figure}

\begin{figure}[t]
    \centering
    \includegraphics[width=0.5\textwidth]{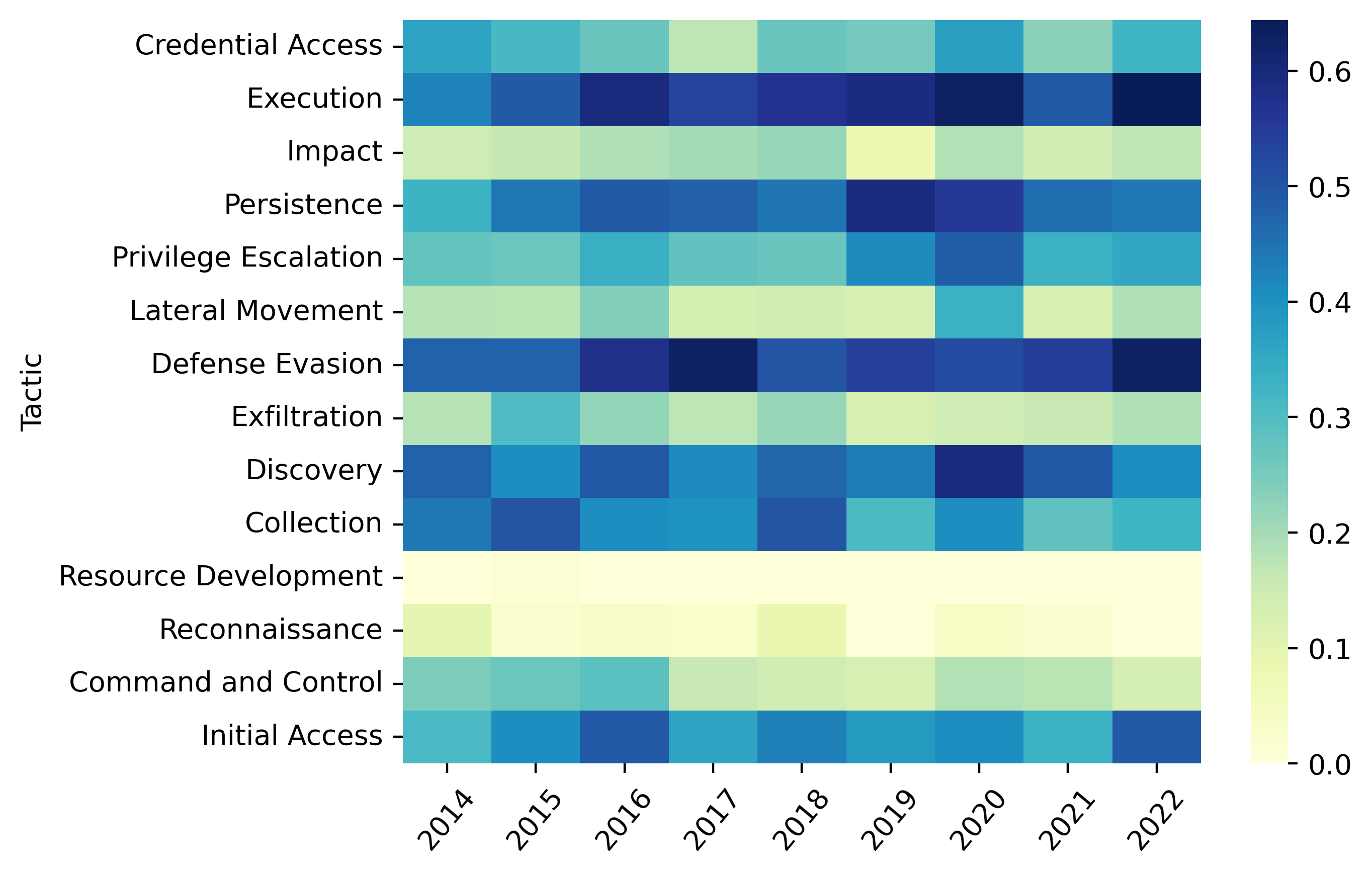}
    \caption{Similarity score of DEFCON offensive talks mapped to the MITRE ATT\&CK Framework from 2014 to 2022.}
    \label{fig:MITRE_years_defcon}
\end{figure}
\begin{figure}[t]
    \centering
    \includegraphics[width=0.5\textwidth]{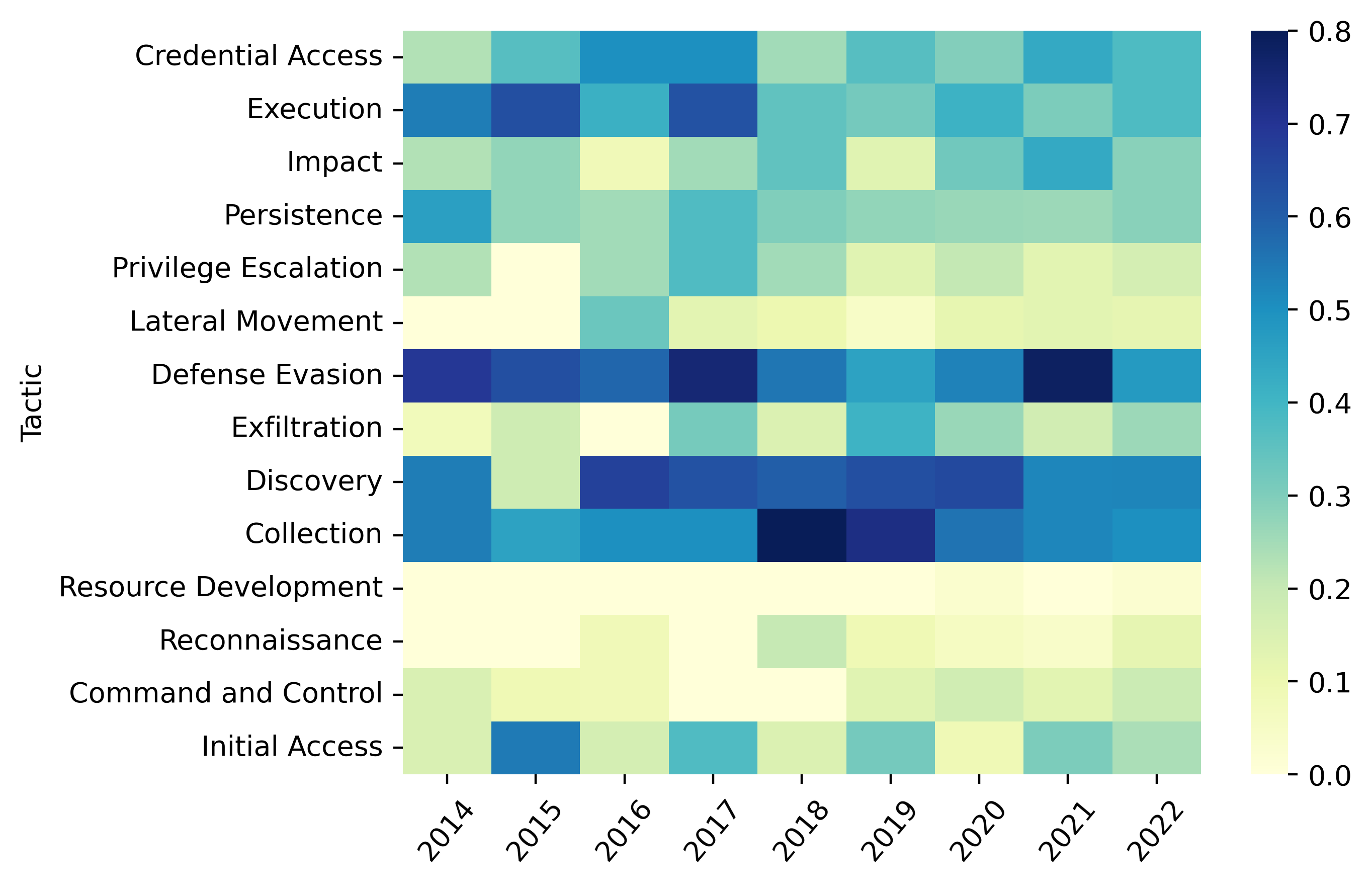}
    \caption{Similarity score of IEEE S\&P offensive talks mapped to the MITRE ATT\&CK Framework from 2014 to 2022.}
    \label{fig:MITRE_years_ieee}
\end{figure}
\begin{figure}[t]
    \centering
    \includegraphics[width=0.5\textwidth]{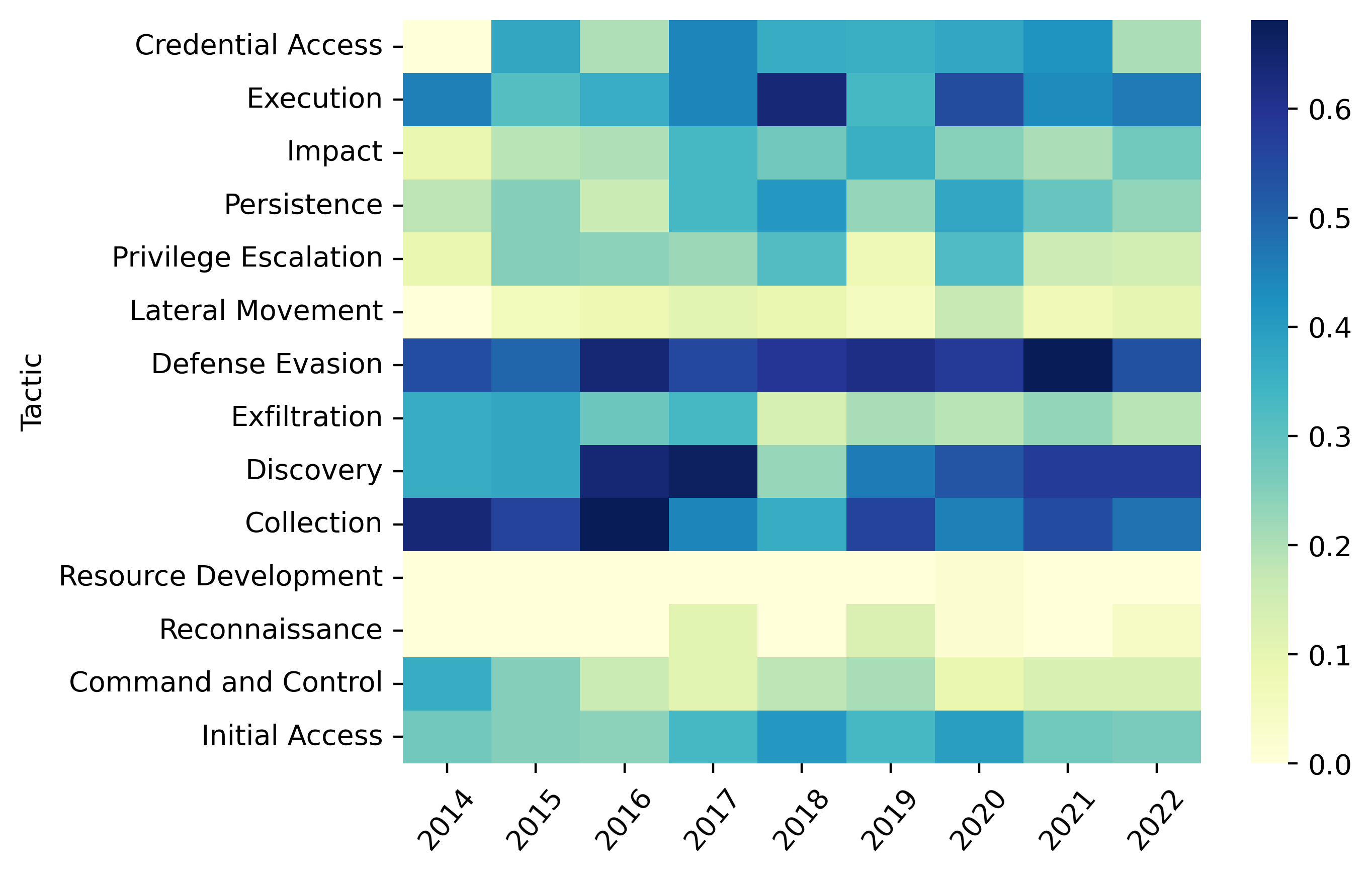}
    \caption{Similarity score of USENIX Security offensive talks mapped to the MITRE ATT\&CK Framework from 2014 to 2022.}
    \label{fig:MITRE_years_usenix}
\end{figure}

\end{document}